\documentclass[final]{IEEEtran}

\usepackage{graphicx}

\usepackage{mathtools}
\usepackage{amsthm,amssymb,graphicx,multirow,amsmath,color,amsfonts}
\usepackage[update,prepend]{epstopdf}
\usepackage[noadjust]{cite}
\usepackage[latin1]{inputenc}
\usepackage{tikz}
\usepackage{bbm} 
\usepackage{pdfpages}
\usepackage{multirow}
\usepackage{subfig}
\usepackage{comment}
\usepackage{amsmath}
\usepackage{nccmath}
\usepackage{hhline}
\newtheorem{rem}{Remark}

\def\nb0{{\mathbf{0}}}
\def\nb1{{\mathbf{1}}}

\newtheorem{lemma}{Lemma}
\newtheorem{thm}{Theorem}

\newtheorem{prop}{Proposition}

\allowdisplaybreaks 

\usepackage{setspace}	

\begin{document}
\graphicspath{{./Figures/}}
\title{
Stochastic differential equations for performance analysis of wireless communication systems
}
\author{
Eya Ben Amar, Nadhir Ben Rached, Ra\'ul Tempone, and Mohamed-Slim Alouini, {\em Fellow, IEEE} \thanks{Eya Ben Amar, Ra\'ul Tempone and Mohamed-Slim Alouini are with King Abdullah University of Science and Technology (KAUST), CEMSE division, Thuwal 23955-6900, Saudi Arabia (email: eya.benamar@kaust.edu.sa; raul.tempone@kaust.edu.sa; slim.alouini@kaust.edu.sa). 

Nadhir Ben Rached is with the School of Mathematics, University of Leeds, Leeds, UK (email: N.BenRached@leeds.ac.uk).

Ra\'ul Tempone is also an Alexander von Humboldt Professor in Mathematics for Uncertainty Quantification, RWTH Aachen University, 52062 Aachen, Germany.
}\vspace{-4mm}}

\maketitle
\begin{abstract}
This paper addresses the difficulty of characterizing the time-varying nature of fading channels. The current time-invariant models often fall short of capturing and tracking these dynamic characteristics. To overcome this limitation, we explore using of stochastic differential equations (SDEs) and Markovian projection to model signal envelope variations, considering scenarios involving Rayleigh, Rice, and Hoyt distributions. Furthermore, it is of practical interest to study the performance of channels modeled by SDEs. In this work, we investigate the fade duration metric, representing the time during which the signal remains below a specified threshold within a fixed time interval. We estimate the complementary cumulative distribution function (CCDF) of the fade duration using Monte Carlo simulations, 
and analyze the influence of system parameters on its behavior. Finally, we leverage importance sampling, a known variance-reduction technique, to estimate the tail of the CCDF efficiently.
\end{abstract}
\begin{IEEEkeywords}
Fade duration, fading channels, importance sampling, Markovian projection, Monte Carlo, stochastic differential equations.
\end{IEEEkeywords}
\section{Introduction}
\subsection{Motivation}
In wireless communication, the propagation of signals from a transmitter to a receiver is often affected by multipath fading. This phenomenon occurs due to the interference and combination of multiple delayed signal paths caused by reflections, scattering, and diffraction in the surrounding environment. Hence, modeling wireless channels is an essential step in the design process. 
Most research in this field employs static models to characterize wireless channels \cite{ takada2008static, avestimehr2011wireless, wang2021research, moraitis2023indoor}. In static models, all channel parameters are assumed to be random but unchanging throughout the observation and estimation phases. However, real-world wireless channels present a dynamic feature. This dynamism develops from various sources, including the mobility of transmitters or receivers. Even when both ends of the communication remain stationary, the propagation environment can undergo temporal changes due to environmental shifts, multipath effects, and interference dynamics. These variations can create fluctuations in signal strength, phase, and signal paths over time, profoundly affecting the reliability and performance of wireless communication systems. 

Notably, stochastic differential equations (SDEs) form a fundamental and robust framework for modeling dynamic systems subjected to random fluctuations.
In particular, they provide an effective structure to capture and analyze the complex behavior of signals in time-varying environments. By incorporating stochastic components, SDEs enable the modeling of wireless channels as dynamic processes with random variations. 
In addition, the performance analysis of wireless communication systems plays a critical role in understanding, evaluating, and optimizing their behavior under various conditions and constraints.
When we consider the temporal dynamics of the signal, this analysis becomes increasingly challenging.

Outage probability is the most commonly known performance metric, which quantifies the probability that the signal falls below a specified threshold, leading to a momentary loss of connectivity. While the outage probability is essential to understanding the reliability of a system, it focuses on the occurrence of outage events without considering their duration. More meaningful second-order statistics of multipath fading channels include the level crossing rate and average fade duration. The level crossing rate indicates how frequently the envelope crosses a specific threshold, whereas the average fade duration measures the duration for which the envelope remains below a given threshold. While these metrics provide valuable insight into the temporal aspects of signal fluctuations and outage occurrences \cite{ dong2001average, beaulieu2003level,cheng2013envelope,issaid2019level}, it is of practical interest to study the distribution of the fading duration rather than only its average value. Previous research has only calculated the average time during which the signal stays below a given threshold, and the density of this duration has not been thoroughly explored.
This work uses dynamical models to study the distribution of the fade duration in a fixed observation interval. 
\subsection{Literature Review}
Modeling wireless channels is an essential step in the design process. Various models exist to describe the statistical characteristics of the multipath fading envelope in wireless communication. The Clark \cite{clarke1968statistical} and Jakes \cite{jakes1994microwave} models are among the earliest and most widely used models for simulating fading channels. The Clark model employs a Rayleigh fading model, if the magnitude of the received signal fluctuates as a random variable (RV) with a Rayleigh distribution. This model is suitable for scenarios where signals traverse multiple paths, experiencing random constructive and destructive interference commonly observed in urban environments or indoors. In contrast, the Jakes model extends the concept of fading by incorporating Doppler shifts caused by the relative motion between the transmitter and receiver.
Although various other models have been proposed \cite{akki1986statistical, akki1994statistical, wang2002double, patel2005simulation}, these studies primarily dealt with deterministic wireless channel models. However, the actual propagation environment continuously varies due to the mobility of nodes at various speeds, resulting in dynamic changes to the network topology. 

Traditional models that assume fixed statistics may no longer suffice to capture the complex dynamics of the propagation environment. For example, the component phases in Clark and Jakes models are assumed to be constant in time. However, in an actual mobile radio communication environment, even without relative motion, the received signal still fluctuates in time. 
Therefore, the traditional Clark and Jakes models are not sufficient to describe such situations. In a stationary (zero Doppler) channel, the Jakes model results in a constant autocorrelation that is inappropriate for long periods due to relative phase fluctuations.

Research has advanced toward stochastic models to address these challenges, offering a better understanding of the dynamic behavior of wireless channels and enhanced modeling accuracy to match real-world conditions \cite{hashemi1993indoor,olama2007recursive, feng2008statistical,charalambous2000general}. For instance, one study \cite{feng2008statistical} extended the traditional Clark model, incorporating the effect of fluctuations in the component phases.

Furthermore, several studies have successfully used SDEs to model multipath fading channels \cite{ olama2009stochastic, charalambous2001state, charalambous2008modeling, charalambous1999stochastic,olama2006stochastic}. 
In particular, the research presented in \cite{olama2009stochastic} introduced a dynamic modeling approach for mobile-to-mobile channels, where system parameters were identified from received signal measurements. Additionally, in \cite{olama2006stochastic}, SDEs were leveraged to model power loss and propose optimal power control algorithms based on the developed models.
The work in \cite{feng2007stochastic} proposed a first-order stochastic autoregressive model for stationary Rayleigh fading channels. The authors developed a two-dimensional (2D) SDE that defines the in-phase and quadrature components of the signal. However, in some scenarios, it becomes more valuable to process a one-dimensional (1D) SDE for the signal envelop. The work in \cite{charalambous1999stochastic} captured the dynamics of the square envelope via a 1D SDE, using Levy's characterization theorem. However, it was established that this model remains valid exclusively under Rayleigh fading conditions. 
\subsection{Contributions}
This work proposes reliable models for time-varying wireless channels using SDEs. More precisely, we employ Markovian projection (MP) to formulate a 1D SDE for the square envelope arising from the 2D SDE for the in-phase and quadrature of the signal. We explore three cases in which the envelope is asymptotically Rayleigh, Rician, or Hoyt. Additionally, we delve into the performance analysis of these dynamic channels by estimating the complementary cumulative distribution function (CCDF) of the fade duration. The problem is equivalent to computing the probability that the fade duration exceeds some limit, and by varying this limit across the observation time interval, we obtain an approximation of the CCDF. In this context, Monte Carlo (MC) simulations are an effective tool to estimate such probabilities.

 In specific scenarios, it becomes rare for the fade duration to exceed a high limit. In practice, it is more desirable for the system to stay above a fixed threshold during the observation interval. Consequently, estimating the tail of the CCDF of the fade duration takes on the role of calculating rare event probabilities. In this context, the crude MC method falls short in estimating such probabilities and necessitates an impractical number of simulations for accurate estimations \cite{kroese2013handbook}. Therefore, it is necessary to use appropriate variance-reduction techniques. Importance sampling (IS) is among the most popular variance-reduction techniques that deliver precise estimates of rare event probabilities with a reduced number of simulations \cite{kroese2013handbook}. We propose an IS estimator to efficiently estimate the right tail of the CCDF of the fade duration.
We also demonstrate that a lower-dimensional SDE for the square envelope is really necessary to obtain an optimal change of measure when IS estimates the right tail of the CCDF of the fade duration.

The paper is structured as follows: Section~\ref{section 1} is dedicated to modeling time-varying wireless channels using SDEs and MP. We assess three cases: Rayleigh, Rice, and Hoyt fading. Section~\ref{Section 2}, introduces the fading duration metric and we proposes an MC estimator to estimate its CCDF. Next, Section ~\ref{Section 3}, presents the motivation for the need for IS and proposes an IS estimator to estimate the tail of the CCDF efficiently.
\section{Stochastic Differential Equations for Modeling Wireless Channels}
\label{section 1} 
\subsection{Problem Setting}
 We employed SDEs to model the random behavior of the signal envelope variations in the time domain. We assumed that the in-phase and quadrature components, denoted as $\{I(s),Q(s)\}_{s \geq 0}$, of the fading signal are Markovian uncorrelated Gaussian random processes with time-varying means and variances. Therefore, $I(s)$ and $Q(s)$ are solutions to the following Ornstein--Uhlenbeck SDEs for $0<t<s<T$, where $t$ is the initial time, and  $T$ is the final time:
\begin{equation}
\label{IQdynamics}
\left\{\begin{aligned}
d I(s) & =k_1 \left(\theta_1-I(s)\right) d s+\beta_1 d W^{(I)}(s), \\
d Q(s) & =k_2 \left(\theta_2-Q(s)\right) d s+\beta_2 d W^{(Q)}(s), \; \; t<s<T \\
I(t) & =I_t \\
Q(t) & =Q_t, \; \; 0<t<T,
\end{aligned}\right.
\end{equation}
where $k_1$, $k_2$, $\theta_1$, $\theta_2$, $\beta_1$, and $\beta_2$ are constants, and $W^{(I)}(s)$ and $W^{(Q)}(s)$ denote independent Wiener processes. From the statistics of $I(s)$ and $Q(s)$, derived in Appendix~\ref{IQ}, $I(s)$ and $Q(s)$ represent Gaussian variables with asymptotic means $\theta_1$ and $\theta_2$ and asymptotic variances $\frac{\beta^2_1}{2 k_1}$ and $\frac{\beta^2_2}{2 k_2}$, respectively.

We let $X(s)=\sqrt{I^2(s)+Q^2(s)}$ be the received signal envelope, where $X(s)$ can have a Rayleigh, Rice, Hoyt, or Beckmann distribution depending on the absence or presence of a strong line of sight \cite{zhu2017distribution}. More precisely, 
\begin{itemize}
    \item If $\theta_1=\theta_2=0$, $k_1=k_2=k$, and $\beta_1=\beta_2=\beta$, then $X(s)$ has an asymptotic Rayleigh distribution with the parameter $\frac{\beta}{\sqrt{2 k}}$.
    \item If $k_1=k_2=k$ and $\beta_1=\beta_2=\beta$, then $X(s)$ has an asymptotic Rice distribution with the parameters ($\sqrt{\theta_1^2+\theta_2^2}$, $\frac{\beta}{\sqrt{2 k}}$).

    \item If $\theta_1=\theta_2=0$, then $X(s)$ has an asymptotic Nakagami-q (Hoyt) distribution \cite{romero2017new} with the shape parameter $q=\frac{\beta_2}{\beta_1} \sqrt{\frac{k_1}{k_2}}$ and $\mathbb{E}[X^2]=\frac{\beta_1^2}{2 k_1}+\frac{\beta_2^2}{2 k_2}$.
    
     \item If $\theta_1\neq \theta_2=0$, $k_1 \neq k_2$, and $\beta_1 \neq \beta_2$, then $X(s)$ has an asymptotic Beckmann distribution.
\end{itemize} 

Based on the model in (\ref{IQdynamics}), the work in \cite{charalambous1999stochastic} has modeled the dynamics of the square envelope $R(s)=I^2(s)+Q^2(s)$, subject to short-term fading, by the following SDE:
\begin{equation}
\label{Rdynamics}
\left\{\begin{aligned}
d R(s) & =\left(c_1-c_2 R(s)\right) d s+c_3 \sqrt{R(s)}d W{(s)}, \; \; t<s<T \\
R(t) & =R_t, \; \; 0<t<T,
\end{aligned}\right.
\end{equation}
where $c_1$, $c_2$, and $c_3$ are constants in $\mathbb{R}$. It used Ito's  differential rule \cite{revuz2013continuous} and Levy's characterization theorem \cite{revuz1991representation}, to transition from a 2D SDE for $I$ and $Q$ to a 1D and Markovian SDE for $R$. The validity of the model in (\ref{Rdynamics}) has been established only in the Rayleigh case. In this work, we employ MP to derive a lower-dimensional, Markovian SDE for $R(s)$ in more general cases, such as when the envelope $X(s)$ is asymptotically Rayleigh, Rician, or Hoyt. 
\subsection{Markovian Projection }
The MP, in the context of SDEs, represents a powerful technique to project a high-dimension SDE into a lower-dimensional one, while preserving the marginal distribution. The following lemma details the dynamics of the projected SDE:
\begin{lemma}{Markovian projection \cite{gyongy1986mimicking}}
\label{MP} \\
We let $X \in \mathbbm{R}^d$ solve 
\begin{equation}
\left\{\begin{aligned}
\mathrm{d} X(t) & =a(t, X(t)) \mathrm{d} t+b(t, X(t)) \mathrm{d} W(t), \quad t \in(0, T] \\
X(0) & =x_0, \quad x_0 \in \mathbb{R}^d,
\end{aligned}\right.
\end{equation}
and we consider the non-Markovian process $S=P X$, where $S \in \mathbbm{R}^{\bar{d}}$ and $P^T=\left(P_1^T, \cdots, P_{\bar{d}}^T\right)^T$ represents the projection matrix onto a dimension $1 \leq \bar{d}<d$. \\ \\
We let $\bar{S}^{\left(x_0\right)} \in \mathbbm{R}^{\bar{d}}$ solve for $t \in[0, T]$
{\small
\begin{equation}
    \left\{\begin{aligned}
d \bar{S}^{\left(x_0\right)}(t) & =\bar{a}^{\left(x_0\right)}\left(t, \bar{S}^{\left(x_0\right)}(t)\right) d t+\bar{b}^{\left(x_0\right)}\left(t, \bar{S}^{\left(x_0\right)}(t)\right) d \bar{W}(t),  \\
\bar{S}^{\left(x_0\right)}(0) & =P x_0,
\end{aligned} \right.
\end{equation}}
where $\bar{W}$ denotes a Wiener process, independent of $W$. The drift and diffusion coefficients of the surrogate process $\bar{S}^{\left(x_0\right)}$ are determined for $1 \leq i, j \leq \bar{d}$ as follows:

{\footnotesize
\begin{equation}
\begin{aligned}
\label{newdriftdiffusion}
\bar{a}^{\left(x_0\right)}(t, s) & =\mathbb{E}\left[P a(t, X(t)) \mid P X(t)=s, X(0)=x_0\right] \\
\left(\bar{b} \bar{b}^T\right)_{i j}^{\left(x_0\right)}(t, s)&=\mathbb{E}\left[\left(P_i^T b b^T P_j\right)(t, X(t)) \mid P X(t)=s, X(0)=x_0\right].
\end{aligned}
\end{equation}} 

Then, $\left.S(t)\right|_{\left\{X(0)=x_0\right\}}=\left.P X(t)\right|_{\left\{X(0)=x_0\right\}}$ and $\bar{S}^{\left(x_0\right)}(t)$ have the same conditional distribution for all $t \in[0, T]$.
\begin{IEEEproof}
See Appendix~\ref{lemmaMP}.
\end{IEEEproof}
\end{lemma} 
In this context, we employed MP by taking $X=(I, Q, R)^T$, $P=(0,0,1)^T$, and $S=R$, where $I$ and $Q$ are solutions of (\ref{IQdynamics}), and $R=I^2+Q^2$. The MP of $R$, denoted as $\Bar{R}$, is obtained by solving the following SDE:
\begin{equation}
\label{MPSDE}
\left\{\begin{array}{rl}
\mathrm{d} \bar{R}(s) & =\bar{a}(s, \bar{R}) \mathrm{d} s+\bar{b}(s,\bar{R}) \mathrm{d} W(s), \quad 0<s<T   \\
\bar{R}(0) & =R(0)\\
\end{array}.\right.
\end{equation}
The determination of the drift $\bar{a}(s, \bar{R})$ and diffusion $\bar{b}(s, \bar{R})$ components relies on the application of MP Lemma~\ref{MP} and depends on the characteristics of the fading environment, whether Rayleigh, Rice, or Hoyt.
\subsubsection{Rayleigh Case}
When the Ornstein--Uhlenbeck  processes $I(s)$ and $Q(s)$ are identically distributed, and do not have a long-term mean adjustment, the SDE (\ref{IQdynamics}) can be expressed as \cite{feng2007stochastic}
\begin{equation}
\label{Rayleigh}
\left\{\begin{aligned}
d I(s) & =-\frac{1}{2} B  I(s) ds + \frac{\sqrt 2}{2} B^{\frac{1}{2}} \sigma dW^{(I)} (s) \quad 0  <s \leq T\\
d Q(s) & =-\frac{1}{2} B  Q(s) ds + \frac{\sqrt 2}{2} B^{\frac{1}{2}} \sigma dW^{(Q)}(s) \quad 0  <s \leq T\\
I(0) & =I_0 \\
Q(0) & =Q_0,
\end{aligned}\right.
\end{equation}
where $B=2 k$ and $\sigma=\frac{\beta}{\sqrt{k}}$. Using Ito's formula, $R(s)$ solves the following non-Markovian SDE: \\
{\footnotesize
\begin{equation}
\left\{\begin{array}{l}
\mathrm{d} R(s)=B\left(\sigma^2 I^2(s)-Q^2(s)\right) \mathrm{d} s +\sqrt{2 B} \sigma[I(s) \quad Q(s)]\left[\begin{array}{c}
\mathrm{d} W^I(s) \\
\mathrm{d} W^Q(s)
\end{array}\right],  \\
R(0)=I^2(0)+Q^2(0)=R_0\\
\end{array}\right.
\end{equation}} \\
From MP Lemma~\ref{MP}, the projected process $\bar{R}$ solves (\ref{MPSDE}) with the projected drift coefficient \\
{\footnotesize
\begin{equation}
\label{deriftrayleigh}
\begin{aligned}
\bar{a}(s, r) & =\mathbb{E}\left[B\left(\sigma^2-I^2(s)-Q^2(s)\right) \mid R(s)=r, I(0)=I_0, Q(0)=Q_0 \right] \\
& =B\left(\sigma^2-r\right),\\
\end{aligned}
\end{equation}}
and the projected diffusion coefficient \\
{\footnotesize
\begin{equation}
\label{diffusionrayleigh}
\begin{aligned}
\bar{b}^2(s,r) & =\mathbb{E}\left[2 B \sigma^2 \left(I^2(s) + Q^2(s) \right) \mid R(s)=r, I(0)=I_0, Q(0)=Q_0 \right] \\
& =2 B \sigma^2 r \\
\Rightarrow \bar{b}(s,r) & =\sigma \sqrt{2 B r}.
\end{aligned}
\end{equation}}
Therefore, for $0<s<T $, (\ref{MPSDE}) can be written as 
{\small
\begin{equation}
\label{dynamicsRbar}
\left\{\begin{array}{rl}
&d \bar{R}(s)=B\left(\sigma^2-\bar{R}(s)\right) d s+\sigma \sqrt{2 B \bar{R}(s)} d W(s),   \\
&\bar{R}(0) =R(0).\\
\end{array}\right.
\end{equation}}
This result precisely corresponds to the SDE form stated in (\ref{Rdynamics}). Consequently, we substantiated the validity of the proposed model, derived using MP rather than Levy's characterization theorem.
\\
We studied the properties of the projected process $\bar{R}$. First, we checked the stationary distribution of $\bar{R}$ using the Fokker--Plank equation. The limit distribution $p(.)$ of the projected process $\bar{R}$ satisfies 
\begin{equation}
\label{FP}
\frac{\partial}{\partial y}(y p(y))=\left(1-\frac{y}{\sigma^2}\right) p(y),  \quad y>0.
\end{equation}
Moreover, the solution of (\ref{FP}) is
\begin{equation}
p(y)=\frac{1}{\sigma^2} e^{\frac{-y}{\sigma^2}}, \quad y>0.
\end{equation}
\normalsize 
Therefore, $\bar{R}(s)$ is an asymptotic exponential distribution with parameter $\frac{1}{\sigma^2}$, which is the exact asymptotic distribution of ${R}(s)$. \\
We also calculated the statistics of $\bar{R}$ in Appendix~\ref{IQ}. The asymptotic means and variances are the same as the statistics of $R$. Furthermore, the expression for the autocorrelation (\ref{autocorr}) is nonconstant in the stationary regime, in contrast to the Clark model. Instead, it exhibits the same exponentially decaying form as the autocorrelation of the fading in the extended Clark model under the condition of a zero Doppler shift, as developed in \cite{feng2008statistical}. \\

\begin{figure}[ht] 
\centering
\includegraphics[scale = 0.4]{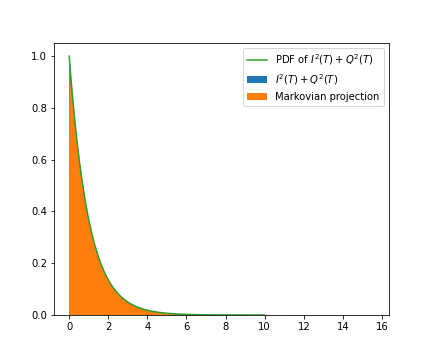} 
\caption {Rayleigh Fading: Histogram of $\bar{R} (T)$ and $I(T)^2+Q(T)^2$ using $10^6$ samples with the following parameters: $T=4$, $N=100$, $I_0=0$, $Q_0=0$, $B=1$, and $\sigma=1$.} 
\label{hist1}
\end{figure}

We solved the original SDE (\ref{IQdynamics}) and the projected SDE (\ref{dynamicsRbar}) and compared in Fig.~\ref{hist1} the histograms of $\bar{R}(T)$ and $I(T)^2+Q(T)^2$. The samples of $\bar{R}(T)$ produce a histogram that closely resembles $I(T)^2+Q(T)^2$.
\vspace{-6mm}

\subsubsection{Rice Fading}
When $I(s)$ and $Q(s)$ have a long-term mean adjustment $\theta$, they solve the following SDE:
\begin{equation}
\label{ricedynamics}
\left\{\begin{aligned}
d I(s) & =k \left(\theta-I(s)\right) d s+\beta d W^{(I)}(s), \\
d Q(s) & =k \left(\theta-Q(s)\right) d s+\beta d W^{(Q)}(s), \; \; 0<s<T \\
I(0) & =I_0\\
Q(0) & =Q_0.
\end{aligned}\right.
\end{equation}
Using Ito's formula, $R(s)$ solves the following non-Markovian SDE, for $s>t$,

 {\small
$$
\left\{\begin{array}{l}
\mathrm{d}R(s)= \left(2 k \theta (I(s)+Q(s)) -2k (I^2(s)+Q^2(s)) +2 \beta^2\right) \mathrm{d} s \\ \quad \quad \quad \quad +\sqrt{2 \beta} \sigma[I(s) \quad Q(s)]\left[\begin{array}{c}
\mathrm{d} W^I(s) \\
\mathrm{d} W^Q(s)
\end{array}\right] \\
R(0)=I^2(0)+Q^2(0)=R_0\\
\end{array}\right.
$$} \\
The projected  process $\bar{R}$ solves (\ref{MPSDE}), with the projected drift 
\begin{equation}
\begin{aligned}
\bar{a}(s, r) & =\mathbb{E}\left[2 k \theta (I(s)+Q(s)) -2k (I^2(s)+Q^2(s))\right. \\ & \left. \quad \quad +2 \beta^2 \mid R(s)=r, I(0)=I_0, Q(0)=Q_0 \right],
\end{aligned}
\end{equation}
and the projected diffusion coefficient 

{\small
\begin{equation}
\begin{aligned}
&\bar{b}^2(s,r) \\ & =4 \mathbb{E}\left[\beta^2 (I^2(s)+ Q^2(s)) \mid R(s)=r, I(0)=I_0, Q(0)=Q_0 \right] \\
& =4 \beta^2 r \\
& \Rightarrow \bar{b}(s,r) =2 \beta \sqrt{r}.
\end{aligned}
\end{equation}}
We let us assume that, initially, $I_0=Q_0$; therefore, $I(s)$ and $Q(s)$ have the same Gaussian distribution at each time $s$, with a mean of
\begin{equation}
    \label{mean}
    m(s)=I_0 \; e^{-k s}+\theta \left(1-e^{-k s}\right)
\end{equation}
and a variance of
\begin{equation}
    \label{variance}
    \sigma^2(s)=\frac{\beta^2}{2 k}\left(1-e^{-2 k s}\right).
\end{equation}
Consequently, 
\begin{equation}
\label{driftrice}
\begin{aligned}
\bar{a}(s, r) & =4 k \theta \; \mathbb{E}\left[ I(s) \mid R(s)=r, I(0)=Q(0)=I_0 \right]\\& -2 k r +2 \beta^2.
\end{aligned}
\end{equation}
To find an explicit expression of the drift $\bar{a}(s, r)$, we computed the conditional expectation $\mathbb{E}\left[ I(s) \mid R(s)=r, I(0)=Q(0)=I_0 \right]$, which reveals how to compute the conditional probability density function (PDF) $f_{I(s) \mid R(s)}(. \mid .)$.
\begin{prop}
\label{prop1}
If $(I(s), Q(s))$ solves (\ref{ricedynamics}), $R(s)=I^2(s)+Q^2(s)$, and $I_0=Q_0$, then, for $ r>x^2$,
    {\footnotesize
\begin{equation}
\label{condpdf}
f_{I(s) \mid R(s)}(x \mid r)=\frac{\exp\left(\frac{x m(s)}{\sigma^2(s)}\right) \left(\frac{r-x^2}{m^2(s)}\right)^{-\frac{1}{4}}\mathcal{
I}_{-\frac{1}{2}}\left( \frac{m(s)\sqrt{ (r-x^2)}}{\sigma^2(s)}\right)}{\sqrt{2 \pi} \sigma (s) \mathcal{
I}_0\left(\frac{m(s) \sqrt{2 r}}{\sigma^2(s)}\right)},
\end{equation}}
where $\mathcal{I}_{\mu}$ is the modified Bessel function of the first kind with order $\mu$. 
\begin{IEEEproof}
See Appendix~\ref{rice}.
\end{IEEEproof}
\end{prop}
Finally,
\begin{equation}
    \label{condexp}
    \mathbb{E}\left[ I(s) \mid R(s)=r \right]=\int x f_{I(s) \mid R(s)}(x \mid r) dx.
\end{equation}
However, if we use the expression (\ref{condexp}) for the conditional expectation to formulate the drift (\ref{driftrice}), solving the SDE (\ref{dynamicsRbar}) becomes computationally expensive because we would need to evaluate the integral in (\ref{condexp}) at each time step and for every sample. Alternatively, we can approximate the conditional expectation (\ref{condexp}) using affine prediction. The conditional expectation can be approximated as follows:
\begin{equation}
\begin{aligned}
 \mathbb{E}[I(s) \mid R(s)=r] \simeq c_1(s)(r-\mathbb{E}[R(s)])+c_2(s),
\end{aligned}
\end{equation}
\\
where $c_1(s)=\frac{\operatorname{Cov}(I(s), R(s))}{\operatorname{Var}(R(s))} $, and $c_2(s)=\mathbb{E}[I(s)]$. Using the properties of the third and fourth moments of Gaussian RVs, we can demonstrate that 

\begin{figure}[ht] 
\begin{center}
\includegraphics[scale = 0.4]{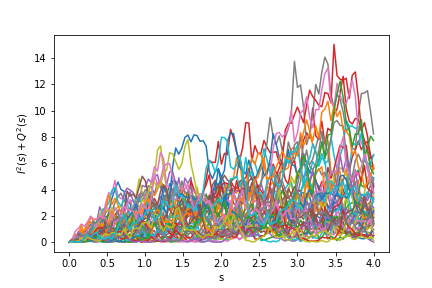} 
\caption {Fifty samples of $I(s)^2+Q(s)^2$ with the following parameters: $N=100$, $I_0=0$, $Q_0=0$, $k=1$, $\theta=1$, and $\beta=1$.} 
\label{figsamples}
\end{center}
\end{figure}

\begin{figure}[ht] 
\begin{center}
\includegraphics[scale = 0.35]{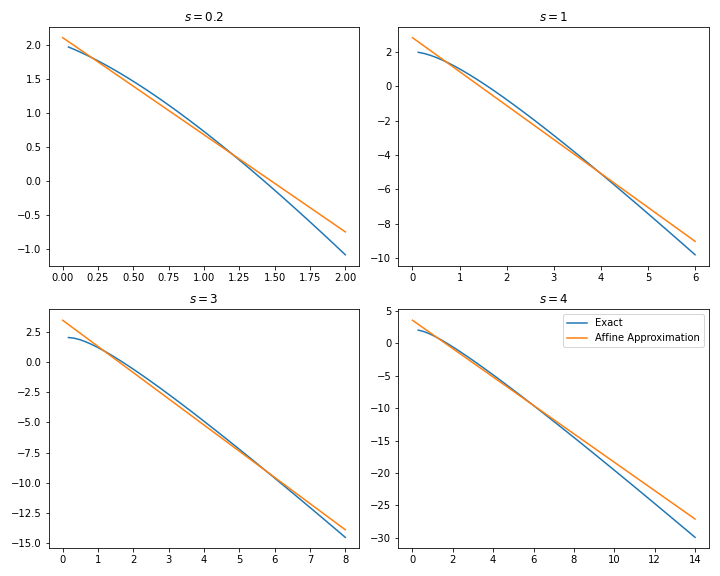} 
\caption {Comparison of the exact and approximated drift with the following parameters: $N=100$, $I_0=0$, $Q_0=0$, $k=1$, $\theta=1$, and $\beta=1$.} 
\label{drift}
\end{center}
\end{figure}

{\small
\begin{equation}
\begin{aligned}
& \operatorname{Cov}(I(s), R(s))=\operatorname{Cov}\left(I(s), I^2(s)+Q^2(s)\right)=2 m(s) \sigma^2(s), \\
\end{aligned}
\end{equation}
\begin{equation}
\begin{aligned}
& \operatorname{Var}(R(s))=\operatorname{Var}\left(I^2(s)+Q^2(s)\right)=4 m^2(s) \sigma^2(s)+2 \sigma^4(s).
\end{aligned}
\end{equation}}
Therefore, 
{\small
\begin{equation}
\begin{aligned}
\label{condexpapprox}
 \mathbb{E}[I(s) \mid R(s)=r] \simeq 
\frac{m(s) \sigma^2(s)\left(r-2\left(\sigma^2(s)+m^2(s)\right)\right)}{4 m^2(s) \sigma^2(s)+2 \sigma^4(s)}+m(s).
\end{aligned}
\end{equation}}
To assess the effectiveness of employing the affine approximation, we compared the exact drift $\bar{a}(s,r)$ obtained by evaluating the exact integral form for the conditional expectation, with the approximated drift $ \tilde{a}(s,r)$ computed using affine prediction. 
Adjusting the range of $r$ based on the fixed time $s$ can yield more meaningful results, as $r$ represents a possible sample of $I^2(s)+Q^2(s)$. Fig.~\ref{figsamples} presents the range of samples for each time, revealing various samples of $I^2(s)+Q^2(s)$ obtained by solving the SDE in (\ref{ricedynamics}). Fig.~\ref{drift} plots $\bar{a}(s, r)$ and $ \tilde{a}(s, r)$ against $r$ for different fixed values of $s$.


Fig.~\ref{drift} reveals that the obtained affine expression is a good approximation for the conditional expectation. Therefore, we can use the linear approximation for the conditional expectation, and  finally, the approximated drift $ \tilde{a}(s,r)$ can be expressed as follows:
\begin{equation}
\label{driftriceapprox}
\begin{aligned}
\tilde{a}(s, r) & \approx 4 k \theta \; \frac{m(s)\sigma^2(s)\left(r-2\left(\sigma^2(s)+m(s)^2\right)\right)}{4 m(s)^2 \sigma^2(s)+2 \sigma^4(s)}\\&+m(s) -2 k r +2 \beta^2,
\end{aligned}
\end{equation}
where $m(s)$ and $\sigma^2(s)$ are given by (\ref{mean}) and (\ref{variance}).\\

\begin{figure}[ht] 
\begin{center}
\includegraphics[scale = 0.4]{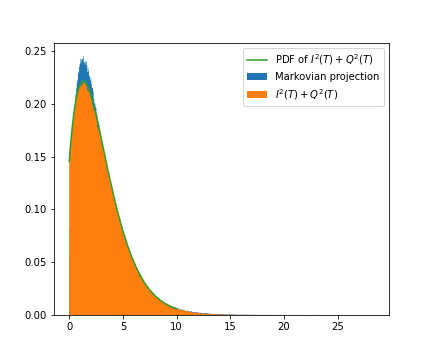} 
\caption {Rice Fading: Histogram of $\bar{R} (T)$ and $I(T)^2+Q(T)^2$ using $10^6$ samples with the following parameters: $N=100$, $I_0=0$, $Q_0=0$, $k=1$, $\theta=1$, and $\beta=1$.} 
\label{histrice}
\end{center}
\end{figure}

To check the efficiency of the proposed projection, in Fig.~\ref{histrice} we plotted the histogram of the solution of the original SDE in (\ref{ricedynamics}) and the histogram of the solution of  projected SDE in (\ref{MPSDE}). The plot demonstrates that the paths of $I^2(T)+Q^2(T)$ and $\bar{R}(T)$ are almost the same; therefore, the approximation is robust. 
\subsubsection{Hoyt Case}
\begin{figure}[ht] 
\begin{center}
\includegraphics[scale = 0.4]{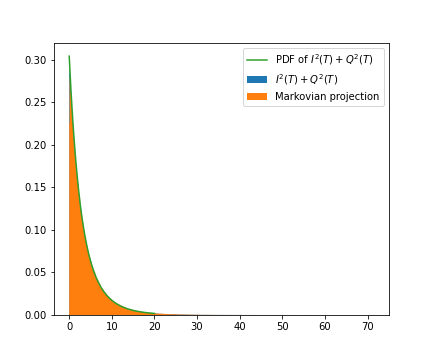} 
\caption {Histogram of $\bar{R} (T)$ and $I(T)^2+Q(T)^2$ using $10^6$ samples with the following parameters: $N=100$, $I_0=0$, $Q_0=0$, $k_1=0.1$, $k_2=0.5$, $\beta_1=\beta_2=1$.} 
\label{histhoyt}
\end{center}
\end{figure}

Another commonly employed statistical distribution for modeling wireless channels is the Hoyt distribution (also known as the Nakagami-q distribution) \cite{romero2017new}. This distribution corresponds to the case in which the in-phase and quadrature are independent jointly Gaussian RVs with a zero mean and different variances. In this case, the processes $I(s)$ and $Q(s)$ satisfy the SDE in (\ref{IQdynamics}), where $\theta_1=\theta_2=0$. By applying Ito's formula to derive the non-Markovian SDE for $R(s)$ and subsequently performing the MP, we can determine the drift and diffusion of the projected SDE in (\ref{MPSDE}) as follows:

{\small
\begin{equation}
\label{drifthoyt}
\begin{aligned}
\bar{a}(s, r)  =&-2 k_1\; \mathbb{E}\left[ I^2(s) \mid R(s)=r, I(0)=I_0, Q(0)=Q_0 \right] +\beta_1^2 \\& -2 k_2\; \mathbb{E}\left[ Q^2(s) \mid R(s)=r, I(0)=I_0, Q(0)=Q_0 \right] +{\beta}_{2}^2,
\end{aligned}
\end{equation}} 
\begin{equation}
\label{diffusionhoyt}
\begin{aligned}
\bar{b}^2(s,r)  &=4 \beta_1^2 \mathbb{E}\left[ I^2(s) \mid R(s)=r, I(0)=I_0, Q(0)=Q_0 \right] \\ & +4 \beta_2^2 \mathbb{E}\left[ Q^2(s) \mid R(s)=r, I(0)=I_0, Q(0)=Q_0 \right].
\end{aligned}
\end{equation} 
We assume that $I_0=Q_0=0$; consequently, $I(s)$ and $Q(s)$ are zero-mean Gaussian distributions with the following variances: 
\begin{equation}
    \label{variance1}
    \sigma^2_1(s)=\frac{\beta_1^2}{2 k_1}\left(1-e^{-2 k_1 s}\right).
\end{equation}
\begin{equation}
    \label{variance2}
    \sigma^2_2(s)=\frac{\beta_2^2}{2 k_2}\left(1-e^{-2 k_2 s}\right).
\end{equation}
To compute the conditional expectation in (\ref{drifthoyt}) and (\ref{diffusionhoyt}), we must determine $f_{I^2(s) \mid R(s)}(\cdot \mid \cdot)$, which is expressed as in Proposition~\ref{prop2}.
\begin{prop}
\label{prop2}
If $(I(s), Q(s))$ solves (\ref{IQdynamics}) with $\theta_1=\theta_2=0$, $R(s)=I^2(s)+Q^2(s)$, and $I_0=Q_0=0$, then, for $r \geq x$,
\begin{equation}
\begin{aligned}
    &f_{I^2(s) \mid R(s)}(x \mid r)\\&=\frac{\left(r-x\right)^{-\frac{1}{2}} \exp\left(-\frac{r-x}{2 \sigma^2_2(s)}-\frac{x}{2 \sigma^2_1(s)}\right) x^{-\frac{1}{2}}}{\pi \exp\left(- \frac{r(\sigma_1^2(s)+\sigma_2^2(s))}{4 \sigma^2_1(s) \sigma^2_2(s)}\right) \mathcal{
I}_0\left(\frac{r (\sigma_1^2(s)-\sigma_2^2(s)) }{4 \sigma^2_1 (s)\sigma^2_2(s)}\right)}
\end{aligned}
\end{equation}    
\begin{IEEEproof}
See Appendix~\ref{hoyt}.
\end{IEEEproof}
\end{prop}
Finally, we can computed the conditional expectation $\mathbb{E}\left[ I^2(s) \mid R(s)=r
\right]$ as follows:
\begin{equation}
    \label{condexp2}
    \mathbb{E}\left[ I^2(s) \mid R(s)=r \right]=\int x f_{I^2(s) \mid R(s)}(x \mid r) dx
\end{equation}
Via the computation, we demonstrate that the simplified expression is given by 
\begin{equation}
\begin{aligned}
 \label{condexp2simp}
    &\mathbb{E}\left[ I^2(s) \mid R(s)=r, I(0)=I_0, Q(0)=Q_0 \right]\\&=\frac{r}{2} \left( 1+\frac{\mathcal{I}_1\left(\frac{r}{4}(\frac{1}{\sigma^2_2(s)}-\frac{1}{\sigma^2_1(s)})\right)}{\mathcal{I}_0\left(\frac{r}{4}(\frac{1}{\sigma^2_2(s)}-\frac{1}{\sigma^2_1(s)})\right)} \right).
\end{aligned}
\end{equation}
Similarly, we can express 
\begin{equation}
\begin{aligned}
    \label{condexp3simp}
    &\mathbb{E}\left[ Q^2(s) \mid R(s)=r, I(0)=I_0, Q(0)=Q_0 \right]\\&=\frac{r}{2} \left( 1+\frac{\mathcal{I}_1\left(\frac{r}{4}(\frac{1}{\sigma^2_1(s)}-\frac{1}{\sigma^2_2(s)})\right)}{\mathcal{I}_0\left(\frac{r}{4}(\frac{1}{\sigma^2_1(s)}-\frac{1}{\sigma^2_2(s)})\right)} \right).
\end{aligned}
\end{equation}
Hence, we have explicit expressions for the drift $\bar{a}(s,r)$ and diffusion $\bar{b}(s,r)$ in the Hoyt case. By using them, we can solve the projected SDE in (\ref{MPSDE}), plot the histogram of the corresponding samples of $\bar{R}(T)$, and compare them to the samples of $I^2(T) + Q^2(T)$, obtained from the original SDE in (\ref{IQdynamics}). The results illustrated in Fig.~\ref{histhoyt} indicate that the MP is an appropriate tool to determine an SDE for the square envelope.
\vspace{-6mm}
\section{Fading Duration}
\label{Section 2}
\subsection{Problem Formulation}
We let $X(s)$ be a random process representing the signal envelope and solve the following SDE for $0<t<s<T$:
\begin{equation}
\label{SDEgeneral}
\left\{\begin{aligned}
d X(s) & =a(s, X(s)) d s+b(s, X(s)) d W(s) \\
X(t) & =x.
\end{aligned}\right.
\end{equation}
We let $Z(T)$ be the time during which $X(s)$ remains below a given threshold $\gamma$, in the time interval $[0,T]$. In other words, $Z(T)$ indicates the fading time of the signal in the period $[0,T]$. Moreover, $Z(T)$ is the solution at final time $T$ of the following ordinary differential equation:
\begin{equation}
\left\{\begin{aligned}
d Z(s) & =\mathbbm{1}_{\{X(s)<\gamma\}} ds, \; \; t<s<T\\
Z(t) & =z, \; \; 0<t<T.
\end{aligned}\right.
\end{equation}
Therefore, 
\begin{equation}
\label{Zexpr}
Z(T)=z +\int_t^T \mathbbm{1}_{\{X(s)<\gamma\}} ds.
\end{equation}
The goal is to determine the distribution of $Z(T)$. To achieve this, we calculated the CCDF of $Z(T)$ given by $P(Z(T)>w| X(t)=x, Z(t)=z)$, for $0 \leq w \leq T$. First, we formulated the problem as a multidimensional SDE:
\begin{equation}
\left\{\begin{aligned}
\label{SDE}
d X(s) & =a(s, X(s)) d s+b(s, X(s)) d W(s) \\
d Z(s) & = c(X(s)) ds , \; \; t<s<T \\
X(t) & =x \\
Z(t) & = {z}, \; \; 0<t<T,
\end{aligned}\right.
\end{equation}
where $c(x)=\mathbbm{1}_{\{x<\gamma\}}$.\\

The quantity of interest, for an initial time $t$ and initial states $x$ and $z$, can be expressed as follows:

\begin{equation}
\label{equ}
\begin{aligned}
&P(Z(T)>w| X(t)=x, Z(t)=z) \\&= \mathbb{E}[g_w( {Z}(T)) \mid X(t)=x, {Z}(t)= {z}],
\end{aligned}
\end{equation}
where $g_w(x)=\mathbbm{1}_{\{x>w\}}$. The distribution of $Z(T)$ is unknown; thus, it is not possible to obtain a closed form of the conditional expectation in (\ref{equ}). Instead, we only have access to samples of $Z(T)$, obtained by numerically solving the SDE in (\ref{SDE}). In this case, one approach to estimating the CCDF is to use MC simulations.

\subsection{Monte Carlo Estimator}
\begin{figure}[ht] 
\begin{center}
\includegraphics[scale = 0.3]{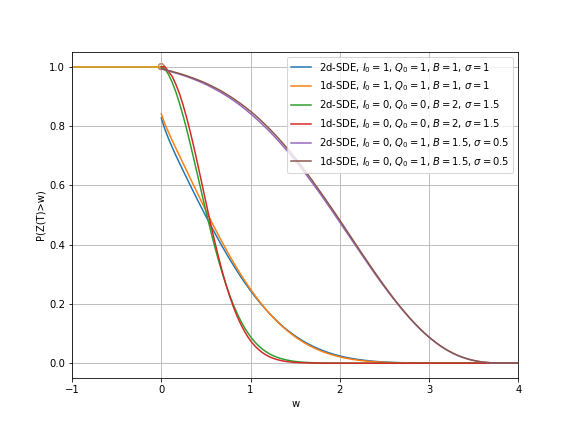} 
\caption {Estimated complementary cumulative distribution function (CCDF) in the Rayleigh case using the Monte Carlo (MC) method with the following parameters: $T=4$, $N=100$, $\gamma=0.5$, and $M=10^6$.} 
\label{ccdfrayleigh}
\end{center}
\end{figure}

\begin{figure}[ht] 
\begin{center}
\includegraphics[scale = 0.3]{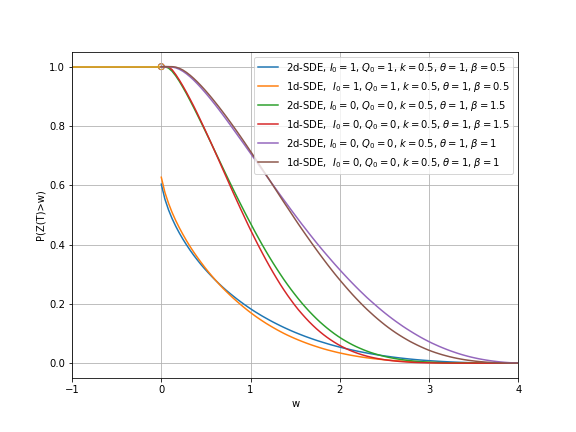} 
\caption {Estimated complementary cumulative distribution function (CCDF) in the Rice case using the Monte Carlo (MC) method with the following parameters: $T=4$, $N=100$, $\gamma=1$, and $M=10^6$.} 
\label{ccdfrice}
\end{center}
\end{figure}

\begin{figure}[ht] 
\begin{center}
\includegraphics[scale = 0.3]{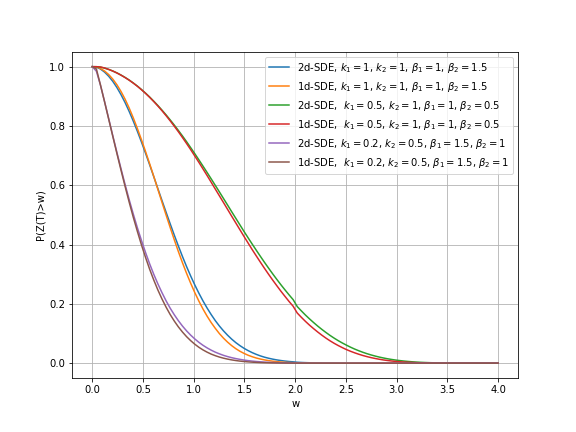} 
\caption {Estimated complementary cumulative distribution function (CCDF) in the Hoyt case using the Monte Carlo (MC) method with the following parameters: $T=4$, $N=200$, $\gamma=0.5$, and $M=10^6$.} 
\label{ccdfhoyt}
\end{center}
\end{figure}

Obtaining analytical solutions to complex SDEs may not always be possible. In practice, numerical methods, such as the Euler--Maruyama scheme, are often used to approximate the solution and simulate the behavior of the system over time. The Euler--Maruyama scheme provides a numerical approximation to the solution of the SDE over a discrete time interval \cite{kloeden1992stochastic}. We consider a time discretization $0=t_0<t_1<\ldots<t_N=T$ and defined $X^N$ and $Z^N$ as the discretized representation of the states $X$ and $Z$, respectively. Given an initial value \(X(t_0)\) at time \(t_0\), the Euler--Maruyama scheme updates the state variables \(X^N(t)\) and \(Z^N(t)\) at each time step \(\Delta t=\frac{T}{N}\) as follows:
\begin{equation}
    \begin{aligned}
    &{X^N}\left(t_0\right)=x_0, \\&
{Z^N}\left(t_0\right)=0,
    \end{aligned}
\end{equation}
and for $n=0, \ldots, N-1,$
{\footnotesize
\begin{equation}
\begin{aligned}
&X^N\left(t_{n+1}\right)=X^N\left(t_n\right) +a\left(t_n,X^N\left(t_n\right) \right)\Delta t +b\left(t_n,X^N\left(t_n\right) \right) \Delta W_n, \\&
Z^N\left(t_{n+1}\right)=Z^N\left(t_n\right) +c(X^N\left(t_n\right)) \Delta t,
\end{aligned}
\end{equation}}
where \(\Delta W_n\) for $n=0,\cdots N-1$, are independent Gaussian RVs $N(0,\Delta t )$. \\ \\
By solving the SDE in (\ref{SDE}) $M$ times, we obtain $M$ independent samples, $\{X^{(k)}(t)\}_{k=1}^M$ and $\{Z^{(k)}(t)\}_{k=1}^M$, of $X(t)$ and $Z(t)$, respectively. Finally,
{\small
\begin{equation}
\mathbb{E}[g_w(Z(T))]\approx \frac{1}{M} \sum_{k=1}^{M} g_w\left({Z}^{(k)}(T) \right)\approx \frac{1}{M} \sum_{k=1}^{M} g_w\left({Z^N}^{(k)}(t_N) \right).
\end{equation}}
This section provides numerical results for estimating the CCDF of the fade duration, denoted as $Z(T)$, in the context of Rayleigh, Rice, and Hoyt fading. We let $I(s)$ and $Q(s)$ solve (\ref{IQdynamics}), and let $\bar{R}(s)$ be the MP of $R(s) = I^2(s) + Q^2(s)$ solving (\ref{MPSDE}). Initially, we employed the MC method to compare the CCDF of $Z(T) = \int_0^T \mathbbm{1}_{\{{I^2(s) + Q^2(s) < \gamma^2}\}} ds$ with the CCDF of $\bar{Z}(T) = \int_0^T \mathbbm{1}_{\{\bar{R}(s) < \gamma^2\}} ds$, aiming to check the effectiveness of working with the lower-dimensional SDE in (\ref{MPSDE}) instead of the original SDE (\ref{IQdynamics}). 
More precisely, the samples of $Z(T)$ are obtained by solving the following SDE:
\begin{equation}
\label{IQdynamicsZ}
\left\{\begin{aligned}
d I(s) & =k_1 \left(\theta_1-I(s)\right) d s+\beta_1 d W^{(I)}(s), \\
d Q(s) & =k_2 \left(\theta_2-Q(s)\right) d s+\beta_2 d W^{(Q)}(s), \; \; 0<s<T, \\
d Z(s) &= f(I(s),Q(s)) ds , \; \; 0<s<T, \\
I(0) & =I_0,\\
Q(0) & =Q_0,\\
Z(0)&=0,
\end{aligned}\right.
\end{equation}
where $f(x,y)=\mathbbm{1}_{\{x^2+y^2<\gamma^2\}}$. The samples of $\bar{Z}(T)$ are obtained by solving the following projected SDE:

\begin{equation}
\label{MPSDEZ}
\left\{\begin{array}{rl}
\mathrm{d} \bar{R}(s) & =\bar{a}(s, \bar{R}) \mathrm{d} s+\bar{b}(s,\bar{R}) \mathrm{d} W(s), \quad 0<s<T,  \\
d \bar{Z}(s) & = \bar{c}(\bar{R}(s)) ds , \; \; 0<s<T, \\
\bar{R}(0) & =I_0^2+Q_0^2, \\
\bar{Z}(0) & =0,
\end{array}.\right.
\end{equation}
where $\bar{c}(x)=\mathbbm{1}_{\{x<\gamma^2\}}$.\\
The projected SDE in (\ref{MPSDEZ}) is obtained using the MP Lemma~\ref{MP}, by taking $X=(I, Q, I^2+Q^2,Z)^T$, $P^T=\left(P_1^T, P_{2}^T\right)$, $P_1=(0,0,1,0)$, $P_2=(0,0,0,1)$ and $S=(I^2+Q^2,Z)^T$. We conducted this comparative analysis across various parameter settings in the three fading cases. The results for the Rayleigh, Rice, and Hoyt cases are presented in Figs.~\ref{ccdfrayleigh}, \ref{ccdfrice}, and \ref{ccdfhoyt}, respectively. 
\vspace{-4mm}
\vspace{-8mm}
\vspace{-8mm}

The remarkable consistency observed across all cases suggests that using the projected process $\bar{R}$ instead of $(I, Q)$, does not noticeably change the CCDF of the fade duration. The discrepancy between the CCDFs of $Z(T)$ and $\bar{Z}(T)$ in the Rice case stems from using the approximated drift $\tilde{a}$ instead of the precise drift $a$. In contrast, for the Rayleigh and Hoyt cases, we employed exact expressions for drift and diffusion. Consequently, the divergence between the CCDFs primarily arises from discretization errors. Significantly, an augmented number of time steps $N$ results in a more accurate fitting of the curves. 

Additionally, in the case where the process initiates above the threshold (by setting $I_0=Q_0=1$), Fig.~\ref{ccdfrayleigh} and~\ref{ccdfrice} reveal an interesting observation: discontinuity occurs at $w=0$. In fact, $P(\bar{Z}(T) = 0)\neq 0$, representing the size of the jump of the represented CCDF. This outcome implies that $\bar{Z}(T)$ is a mixed RV with a jump at zero. This phenomenon is explained by the fact that there is a non-zero probability that the signal will remain above the threshold for the entire interval $[0, T]$ when the process starts above the threshold and with specific parameters.

In the next experiment, we employed the MC method to examine how the distribution of the fade duration $Z(T)$ behaves when we manipulate the parameters of the channel. Specifically, in the Rayleigh case, the parameters $B$ and $\sigma$ exert control over the autocorrelation of the complex envelope $\Psi=I+iQ$. Notably, $I(t)$ and $Q(t)$ are independent; thus, we can express the autocorrelation of $\Psi$ as $C_{\psi \psi}(\Delta t)= C_{II}(\Delta t) + C_{QQ}(\Delta t) \stackrel{T \rightarrow \infty}{=} \sigma^2 e^{-\frac{1}{2} B \Delta t}$. Consequently, $B$ represents the correlation time scale, and the autocorrelation of the signal increases linearly with $\sigma^2$.

\begin{figure}[ht]
\centering
    \hspace{-0.8cm}
  \captionsetup{justification=centering}
\subfloat[$\sigma=1$]{\includegraphics[width=1.9in]{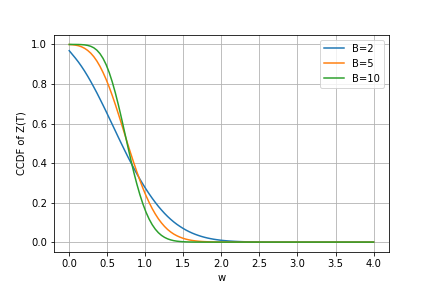}}
\captionsetup{justification=centering}
\hspace{-0.7cm}
\subfloat[$B=2$]{\includegraphics[width=1.9in]{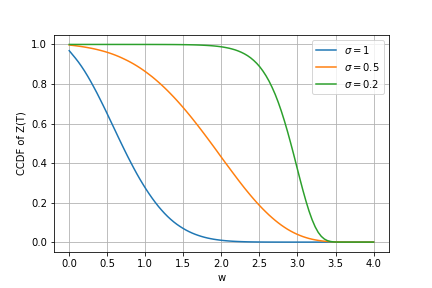}}
\hspace{-0.5cm}
  \caption {Estimated complementary cumulative distribution function (CCDF) using the Monte Carlo (MC) method, for various values of $B$ and $\sigma$ with the following fixed parameters: $T=4$, $N=100$, $I_0=1$, $Q_0=1$, $\gamma= 0.5$, and $M=10^6$.} 
\label{compBsigma}
\end{figure}

Fig.~\ref{compBsigma} presents a comparative analysis of the behavior of the estimated CCDF of $Z(T)$ in the Rayleigh case by solving (\ref{IQdynamicsZ}), as we manipulated the parameters $B$ and $\sigma$. Altering the parameter $B$ has a relatively minor effect on the shape of the CCDF and the associated calculated statistics. In contrast, when the parameter $\sigma$ is adjusted, even a slight modification can result in a significant transformation in the distribution of $Z(T)$. Furthermore, the results demonstrate that as both $B$ and $\sigma$ increase, the CCDF experiences a more rapid decay towards zero. Additionally, as $B$ increases and $\sigma$ decreases, the probability $P(Z(T)=0)$ tends toward zero, signifying that the RV $Z(T)$ transitions towards a continuous distribution when the signal exhibits lower levels of correlation. 
\vspace{-4mm}

\begin{figure}[ht] 
\begin{center}
\includegraphics[scale = 0.4]{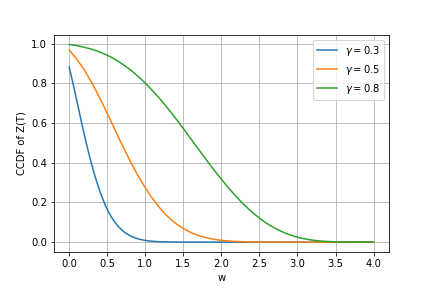} 
\caption {Estimated complementary cumulative distribution function (CCDF) using Monte Carlo (MC) method, for various values of $\gamma$ with the following parameters: $T=4$, $N=100$, $I_0=1$, $Q_0=1$, $B=2$, and $\sigma=1$.} 
\label{compth}
\end{center}
\end{figure}

Another noteworthy experiment involves keeping the channel parameters fixed while varying the threshold $\gamma$. Fig.~\ref{compth} presents the variations of the CCDF of $Z(T)$. The observations indicate that as the threshold rises, the CCDF of $Z(T)$ undergoes a more rapid decay toward zero, and the magnitude of the jump at zero diminishes.
\vspace{-4mm}
\section{Importance Sampling for Right-Tail Estimation}
\label{Section 3}
\subsection{Motivation}
\begin{figure}[ht] 
\begin{center}
\includegraphics[scale = 0.4]{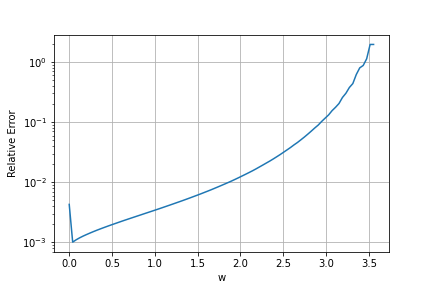} 
\caption {Relative error of the Monte Carlo (MC) estimator of $P(Z(T)>w)$ with the following parameters  $N=100$, $I_0=1$, $Q_0=1$, $B=1$, $\sigma=1$, and $\gamma=0.5$.} 
\label{motivation}
\end{center}
\end{figure}

When estimating $P(Z(T)>w)$, it becomes evident that the MC method tends to estimate a probability of zero when $w$ approaches $T$ from the left. However, we anticipated that the actual value is extremely small but not precisely zero given that the occurrence of $\{Z(T)>w\}$ is rare when $T-w$ is small. Moreover, the MC estimations tend to be inaccurate for such infrequent events. To illustrate this, we calculated the relative error, defined using the central limit theorem as follows:
 \begin{equation}
     \label{eqn:eq10}
\epsilon_{MC}=C \frac{\sqrt{P(Z(T)>w)(1-P(Z(T)>w))}}{\sqrt{M_{MC}} P(Z(T)>w)},
\end{equation}
where $C$ denotes the confidence constant equal to 1.96 (for the 95 \%
confidence interval), and $M_{MC}$ 
represents the number of samples for the MC estimator. Fig.~\ref{motivation} displays the relative error within the range for $w$, using $M_{MC}=10^6$.
\vspace{-4mm}

The relative error significantly increases approaching the right tail. This phenomenon implies that the MC method requires more samples to achieve a given relative error as the value of $w$ increases. Notably, even with a sample size of $10^6$, MC struggles to estimate $P(Z(T)>w)$ accurately for $w>3.5$. Consequently, we must employ variance-reduction techniques to minimize the required number of samples for estimating the tail of the CCDF. We employed IS as one such technique.

\subsection{Importance Sampling}
We present the idea of IS in the general case before applying it to SDEs. We consider the estimation of $\mathbb{E}[g(X)]$, where $X$ is an RV with PDF $f(.)$, and $g$ is a given function. In scenarios where $g$ takes the form of an indicator function, signifying probability estimation, the region of rare events occurs when this quantity becomes extremely small. In this case, standard MC simulations require considerable samples to obtain accurate estimates. In addition, the IS method is among the most popular variance-reduction techniques used to overcome the failure of naive MC simulations and reduce the computational work considerably \cite{rached2015unified, ben2021efficient, ben2023state}. The idea is to modify the sampling distribution so that the rare event is generated with a higher probability than under the
original distribution \cite{kroese2013handbook}. We consider a new PDF $\tilde{f}$ such that
\begin{equation}
 supp \Big  \{ f(.) g(.) \Big \} \; \; \subset \; \; {supp \{ \tilde{f}(\cdot)} \},
 \end{equation}
where ${supp}\{ f \}=\{x \in \mathbb{R}^d \mid f(x ) \neq 0\}$. The IS technique consists of writing the quantity of interest as follows
\begin{equation}
\begin{aligned}
\mathbb{E}[g(X)] &=\int_{\mathbb{R}^d} g(x) f(x) d x \\
&=\int_{\mathbb{R}^d} g(x) \frac{f(x)} {\tilde{f}(x)} \tilde{f}(x) d x \\&=\mathbb{E}_{\tilde{f}}\left[\tilde{g}\left(X\right)\right],
\end{aligned}
\end {equation}
where 
\begin{equation}
\tilde{g}\left(x\right)=g\left(x\right) \frac{f\left(x\right)}{\tilde{f}\left(x\right)},
\end{equation}
and $\mathbb{E}_{\tilde{f}}$ is the expectation under which the RV $X$ has the PDF $\tilde{f}(\cdot)$.
\\ \\
Then, the IS estimator is  
\begin{equation}
\mathcal{A}_{I S}=\frac{1}{M} \sum_{k=1}^{M} \tilde{g}\left(X^{(k)}\right),
\end{equation}
where $\{X^{(k)} \}_{k=1}^{M}$ represents independent realizations of $ X$ sampled according to $\tilde{f}(\cdot)$.

In the context of SDEs, if $X(s) \in \mathbb{R}^d$ solves the following SDE: 
\begin{equation}
\label{SDEgd}
\left\{\begin{aligned}
d X(s) & =a(s, X(s)) d s+b(s, X(s)) d W(s), 0<s<T \\
X(0) & =x_0,
\end{aligned}\right.
\end{equation}
and we aim to estimate $\mathbb{E}\left[g\left(X(T)\right)\right]$, the change of measure in the IS method is performed in the path space induced by the Wiener process. We applied a change of measure to the SDE in (\ref{SDEgd}) such that we minimized the variance of the MC estimator. We formulated the problem of finding an optimal change of measure as a stochastic optimal control problem. To better understand this, we let $X^N(t)$ be the forward Euler--Maruyama approximation of the proposed process $X(s)$. Then, 
\begin{equation}
\left\{\begin{array}{l}
X^N(t_{n+1}) = X^N(t_n) + a(X^N(t_n), t_n) \Delta t \\ \quad \quad  \quad  \quad \quad + b(X^N(t_n), t_n) \sqrt{\Delta t}\epsilon_n, \quad  n=0, \ldots, N-1 \\
X^N\left(t_0\right)=X(0)=x_0,
\end{array}\right.
\end{equation}
where $\epsilon_n, n=1, \cdots, N$ are independent and identically distributed standard normal vectors. We let $\zeta_n=\zeta(X^N(t_n), t_n)$ be the control at the time step $t_n$ and we propose a measure change on each $\epsilon_n$, of the following form: 
\begin{equation}
\hat{\epsilon}_n=\sqrt{\Delta t} \zeta_n+\epsilon_n.
\end{equation}
Therefore, the new path $X^N_{\zeta}$ satisfies
{\footnotesize
\begin{equation}
\begin{aligned}
\left\{\begin{array}{l}
X^N_{\zeta}(t_{n+1}) = X^N(t_n) +\left( a(X^N_{\zeta}(t_n), t_n)+ b(X^N_{\zeta}(t_n), t_n) \zeta_n\right)  \Delta t \\ \quad \quad  \quad  \quad \quad + b(X^N_{\zeta}(t_n), t_n) \sqrt{\Delta t}\epsilon_n\\
X^N_{\zeta}(t_{0})=X^N\left(t_0\right)=x_0.
\end{array}\right.
\end{aligned}
\end{equation}}
The objective is to minimize the variance of the resulting estimator. As the IS estimator is unbiased, minimizing the variance is equivalent to minimizing the second moment. To maintain unbiasedness, we introduced the likelihood as follows:
\begin{equation}
L_n\left(\hat{\epsilon}_n\right)=\exp \left\{-\frac{1}{2}\left\|\hat{\epsilon}_n\right\|^2\right\} \exp \left\{\frac{1}{2} \| \hat{\epsilon}_n-\sqrt{\Delta t} \zeta_n\|^2\right\},
\end{equation}
After replacing the expression of $\hat{\epsilon}_n$, the likelihood function becomes
\begin{equation}
L_n\left(\epsilon_n\right)=\exp \left\{-\frac{1}{2} \Delta t\left\|\zeta_n\right\|^2-\sqrt{\Delta t}\left\langle\epsilon_n, \zeta_n\right\rangle\right\},
\end{equation}
where $\langle\cdot, \cdot\rangle$ is the dot product between two vectors, and $||.||$ is the Euclidean norm of the vector.

Finally, the quantity of interest can be expressed as follows:
\begin{equation}
\mathbb{E}\left[g\left(X^N(T)\right)\right]=\mathbb{E}\left[g\left(X_\zeta^N(T)\right) \prod_{n=0}^{N-1} L_n\left(\epsilon_n\right)\right].
\end{equation}
The objective is to minimize the second moment; thus, the cost function is defined as follows:
{\footnotesize
\begin{equation}
\begin{aligned}
&C_{t, x}(\zeta_0, \cdots, \zeta_{N-1})=
\mathbb{E}\left[g^2\left(X_\zeta^N(T)\right) \prod_{n=0}^{N-1} L_n^2\left(\epsilon_n\right) \mid X_\zeta^N(t)=x\right] \\
=&\mathbb{E}\left[g^2\left(X_\zeta^N(T)\right) \right. \\& \left. \times \prod_{n=0}^{N-1} \exp \left\{-\Delta t\left\|\zeta_n\right\|^2-2\left\langle\epsilon_n, \zeta_n\right\rangle \sqrt{\Delta t}\right\} \mid X_\zeta^N(t)=x\right].
\end{aligned}
\end{equation}}
Taking the limit as $\Delta t \rightarrow 0$ leads to the following:
{\small 
\begin{equation}
\begin{aligned}
C_{t, x}(\zeta)=\mathbb{E}\left[\right.&g^2\left(X_\zeta(T)\right) \exp \left\{\right.-\int_t^T\left\|\zeta\left(s, X_\zeta(s)\right)\right\|^2 \mathrm{~d} s   \\ & - 2 \int_t^T\left\langle\zeta\left(s, X_\zeta(s)\right), \mathrm{d} W(s)\right\rangle \left. \right\} \mid X_\zeta(t)=x \left. \right],
\end{aligned}
\end{equation}}
where $X_\zeta$ is subject to the controlled SDE dynamics \\
\begin{equation}
\label{SDEIS}
\left\{\begin{array}{l}
\mathrm{d} X_\zeta(s)=\left(a\left(s, X_\zeta(s)\right)+b \left(s, X_\zeta(s)\right) \zeta\left(s, X_\zeta(s)\right)\right) \mathrm{d} s \\ \quad \quad  \quad  \quad +b \left(s, X_\zeta(s)\right) \mathrm{d} W(s), \quad 0<s<T \\
X_\zeta(t)=X(t)=x.
\end{array}\right.
\end{equation}
We defined the value function as follows:
\begin{equation}
\label{valuefun}    
u(t, x)=\min _{\zeta \in \mathcal{B}} C_{t, x}(\zeta),
\end{equation}
where $\mathcal{B}$ is the set of admissible Markov controls. 
The parameter $\zeta$ serves as the IS parameter, which is responsible for specifying the change of measure. One conceivable approach to identifying a suitable control $\zeta$ involves the application of a parametric family of change of measure and solving the optimization problem in (\ref{valuefun}) to determine the optimal parameters that provide the optimal control $\zeta$ within the preselected class of measures. In contrast, if the aim is to determine the optimal control in the full spectrum of possible Markov controls, it is best to apply Theorem~1 \cite{nusken2021solving}.
\begin{thm}
\label{thm1}
We assume that the value function $u(t, x)$ defined in (\ref{valuefun}) is bounded and smooth and let $D$ represent the support of $u$, meaning that $u(t, x)$ is nonzero for $(t, x)\in D$. Then, $u(t, x)$ solves the following partial differential equation (PDE):
\begin{equation}
\begin{aligned}
\label{pde1}
\left\{\begin{array}{l}
\partial_t u+  \sum_{i=1}^d   a_i\partial_{x_i} u+\frac{1}{2} \sum_{i,j,k=1}^d  b_{ik} b_{jk} \partial_{x_i x_j} u \\   -\frac{1}{4 u} \sum_{i=1}^d \left(\sum_{j=1}^d \partial_{x_j}u \; b_{ij}  \right)^2 =0, \quad(t, x) \in D \\
u(T, .)=g^2,
\end{array}\right.
\end{aligned}
\end{equation}
where $a_i$ denotes the $i$th component of the drift vector $a$, $b_{ij}$ represents the element in the $i$th row and $j$th column of matrix $b$, $\partial_{x_i} u$ denotes the derivative of $u$ with respect to the $i$th component of vector $x$, and $\partial_{x_i x_j} u$ is the second derivative of $u$ with respect to the $i$th and $j$th components of vector $x$. The optimal control is computed as follows:
\begin{equation}
\label{OC}
\zeta^*(t, x)=\frac{1}{2} b(t, x) \nabla \log u(t, x),
\end{equation}
where $\nabla$ denotes the gradient vector.
\begin{IEEEproof}
See the proof in \cite{rached2022single}.
\end{IEEEproof}
\end{thm}

\begin{rem}
We assume that $g$ does not change the sign. If we introduce the change of variable $u(t, x) = v^2(t,x)$ and substitute $u(t, x)$ with $v^2(t,x)$ in equation (\ref{pde1}), we can deduce that $v$ exactly solves the Kolmogorov backward equations (KBEs) given by
\begin{equation}
\label{kbe}
\left\{\begin{array}{l}
\partial_t v+  \sum_{i=1}^d   a_i\partial_{x_i} v+\frac{1}{2} \sum_{i,j,k=1}^d  b_{ik} b_{jk} \partial_{x_i x_j} v =0, \; t<T,\\
v(T, .)=g.
\end{array}\right.
\end{equation}
\end{rem}
The KBE formulates a deterministic PDE whose solution may be written in the form of an expectation value of a terminal condition \cite{oksendal2003stochastic}. The solution of the KBE (\ref{kbe}) corresponds to the conditional expectation $\mathbb{E}[g(X(T)) \mid X(t) = x]$. The quantity $u(t, x)$ defined in (\ref{valuefun}) represents the second moment of the IS estimator. Given that $v(t, x)$ precisely represents the quantity to estimate, by employing the optimal control (\ref{OC}), the second moment of the IS estimator equals the square of the first moment, resulting in a variance of zero.
\begin{rem}
Solving the KBE in (\ref{kbe}) suffers from the curse of dimensionality, wherein the computational cost escalates exponentially as the dimensionality $d$ increases.
\end{rem}
\subsection{Numerical Results}
This section proposes an IS estimator for the right-tail estimation of the CCDF of the fade duration in the  Rayleigh case. We determined an optimal change of measure, as described in Theorem~\ref{thm1}. Obtaining the optimal change of measure entails solving the KBE in (\ref{kbe}). Applying the change of measure on the SDE in (\ref{IQdynamicsZ}) necessitates solving a PDE with three spatial dimensions and one temporal dimension, which can be computationally expensive and, given limited resources, may even become infeasible. To circumvent the challenges posed by the curse of dimensionality inherent in solving the corresponding PDE, we applied the change of measure on the projected SDE in (\ref{MPSDEZ}), where the drift $\bar{a}(s, r)$ is determined by (\ref{deriftrayleigh}) and the diffusion $\bar{b}(s, r)$ is specified by (\ref{diffusionrayleigh}). 

\begin{figure}[ht]
\centering
    \hspace{-0.8cm}
  \captionsetup{justification=centering}
\subfloat{\includegraphics[width=1.9in]{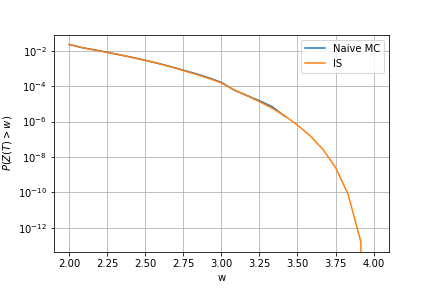}}
\captionsetup{justification=centering}
\hspace{-0.7cm}
\subfloat{\includegraphics[width=1.9in]{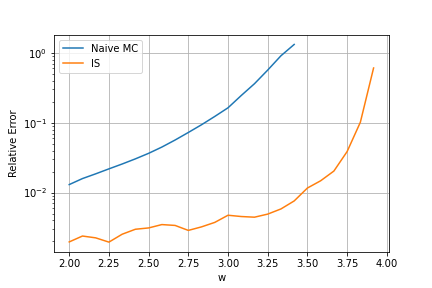}}
\hspace{-0.5cm}
  \caption {Estimated tail of the complementary cumulative distribution function (CCDF) and the corresponding relative error using the Monte Carlo (MC) and importance sampling (IS) methods with the following parameters: $T=4$, $N=100$, $I_0=1$, $Q_0=1$, $B=1$, $\sigma=1$, $\gamma=0.5$, and $M_{MC}=M_{IS}= 10^6$.} 
\label{ISRelErr}
\end{figure}

This approach underscores the significance of identifying a lower-dimensional SDE for modeling fading channels, as opposed to a potentially problematic higher-dimensional SDE in specific scenarios. The modified controlled SDE takes the following form:

{\footnotesize
\begin{equation}
\label{dynamicsRbarxiIS}
\left\{\begin{array}{rl}
&d \bar{R}_{\zeta^*}(s)=B\left(\sigma^2-\bar{R}_{\zeta^*}(s)\right) d s \\ & +\sigma \sqrt{2 B \bar{R}_{\zeta^*}(s)} \; \bar{\zeta}^*(s,\bar{R}_{\zeta^*}(s),\bar{Z}_{\zeta^*}(s))
d s+\sigma \sqrt{2 B \bar{R}_{\zeta^*}(s)} d W(s), \\
& d \bar{Z}_{\zeta^*}(s) = \bar{c}(\bar{R}_{\zeta^*}(s)) ds ,\\
&\bar{R}_{\zeta^*}(0) =\bar{R}(0),\\
&\bar{Z}_{\zeta^*}(0) =\bar{Z}(0),\\
\end{array}.\right.
\end{equation}}
where the optimal control is defined as 
\begin{equation}
\label{OC2}
\zeta^*(t, x,z)=\sigma \sqrt{2 B x} \; \partial_x \log v(t, x,z),
\end{equation}
and $v(t,x,z)=\mathbb{E}[g_w(\bar{Z}(T)) \mid \bar{R}(t) = x, \bar{Z}(t)=z]$ is the solution of the following PDE, for $x \geq 0,\; 0 \leq t\leq T, \; \text{and} \;  t-(T-w)\leq z \leq t$,
\begin{equation}
\label{PDE}
\left\{\begin{aligned}
&\partial_t {v}(t,x,z)+B\left(\sigma^2-x\right) \partial_x {v}(t,x,z)+\bar{c}(z) \partial_{z} {v}(t,x,z) \\ &+\sigma^2 B x \partial_{x x}{v}(t,x,z)=0, \\
&{v}(T, x,z)=g_w(z), \quad \text{(+ Boundary conditions)}.
\end{aligned}\right.
\end{equation}
The PDE solver and the boundary conditions of (\ref{PDE}) are derived in Appendix~\ref{KBEsolver}. The optimal IS estimator is given by 
\begin{equation}
\mathcal{A}_{I S^*}=\frac{1}{M_{IS}} \sum_{k=1}^{M_{IS}} g_w \left(\bar{Z}_{\zeta^*}^{(k)}\right),
\end{equation}
where $M_{IS}$ is the number of samples for the IS estimator. Moreover, the relative error of the IS estimator can be expressed as follows 
 \begin{equation}
 \epsilon_{I S}=C \frac{\sqrt{\operatorname{Var}\left[\mathcal{A}_{I S^*}\right]}}{\sqrt{M_{I S}} P(Z(T)>w)}.
\end{equation}
We evaluated the unbiasedness of the IS estimators by generating Fig.~\ref{ISRelErr}, displaying the estimated CCDF for values of $w$ greater than two and the corresponding relative error. \\

\begin{figure}[ht] 
\begin{center}
\includegraphics[scale = 0.45]{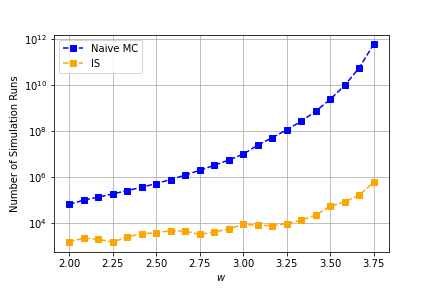} 
\caption {Number of required simulation runs for a 5\% relative error for the projected process, using the optimal IS and naive MC methods, with the following parameters: $T=4$, $N=100$, $I_0=1$, $Q_0=1$, $B=1$, $\sigma=1$, and $\gamma=0.5$.} 
\label{comparison}
\end{center}
\end{figure}

We observed consistency in the estimation of the CCDF using the MC and IS methods for $w<3.5$. The MC estimator estimates $P(Z(T)>w)$ at zero for $w\geq 3.5$ but exhibits a remarkably high relative error. Conversely, the IS estimator approximates the right tail of the CCDF at minimal values with a reasonable relative error. Fig.~\ref{comparison} presents the number of samples needed to achieve a 5\% relative error using the IS and MC methods.

A substantial amount of variance reduction occurs when using IS compared to the crude MC method. For a better view of the results, Table~\ref{tabw} presents the estimated value of $P(Z(T)>w)$ and corresponding variance for various values of $w$, using both estimators. We employed $10^6$ samples for the MC and IS estimators across the entire range for $w$. \\

{\footnotesize
\begin{table}[h!]
	\centering
	\begin{tabular}{ | c|| c| c|| c|c|c|c| c|}
		\hline
		 $w$& $\mathcal{A}_{MC}$ &$\mathcal{A}_{I S^*}$& $Var[\mathcal{A}_{MC}]$& $Var[\mathcal{A}_{I S^*}]$ \\
		\hline
		$2.5$ & $0.003$& $0.0028$& $0.0034 $&$2.09 \times 10^{-5}$ \\
		\hline
		$3$ & $1.8 \times 10^{-4}$ & $1.5 \times 10^{-4}$ &$1.8\times 10^{-4}$&$1.16 \times 10^{-7}$\\
		\hline
  		$3.25$ & $1.7 \times 10^{-5}$& $1.34 \times 10^{-5}$&$1.7 \times 10^{-5}$&$1.11 \times 10^{-9}$\\
		\hline
		$3.5$& $0$ & $6.048 \times 10^{-7}$& &$1.41 \times 10^{-11}$\\
  		\hline
		$3.75$ & 0&$2.58 \times 10^{-9}$ & &  $2.59 \times 10^{-15}$\\
		\hline
  		$3.83$ & 0&$7.5 \times 10^{-11}$ & & $1.51 \times 10^{-17}$\\
		\hline
	\end{tabular}
 	\caption{Comparison of the Monte Carlo (MC) and importance sampling (IS) estimators for values of $w$ with the following parameters: $T=4$, $N=100$, $I_0=1$, $Q_0=1$, $B=1$, $\sigma=1$, and $\gamma=0.5$.}
	\label{tabw}
\end{table}}
Table~\ref{tabw} reveals that the optimal IS estimator effectively estimates very small values, as low as $10^{-9}$, with a relative error below $5 \%$. In addition, it estimates $P(Z(T) > 3.83)$ as $7.5 \times 10^{-11}$ with only a $10 \%$ relative error. 
\section{Conclusion}
This work developed more realistic stochastic models to represent time-varying wireless channels, using SDEs and MP to investigate channel envelope dynamics across various fading scenarios. The results for the Rayleigh case align with those developed using a different method in \cite{charalambous1999stochastic}. Furthermore, we extended this analysis to derive SDEs for more general cases, including the Rice and Hoyt envelopes.

The paper also focused on the performance analysis of stochastic channels. Notably, we introduced the fading duration metric, which measures the time during which the channel remains faded (i.e., below a given threshold). Using the dynamical models derived in the first part, we estimated the CCDF of this duration in the Rayleigh, Rice, and Hoyt cases. We employed the MC method on the original and projected SDEs and demonstrated the consistency of the results, verifying the effectiveness of using MP. Additionally, we explored how the CCDF behaves when altering the signal autocorrelation parameters.

Finally, we used IS to estimate the tail of the CCDF accurately. We proposed an optimal change of measure for the Rayleigh case, highlighting significant gains in terms of variance reduction, yielding more precise tail estimations.
\appendices
\section{Statistics of $I(T)$, $Q(T)$ and $\bar{R}(T)$}\label{IQ}
From the SDE in (\ref{IQdynamics}), the expression of $I(T)$ can be written as follows:
{\footnotesize
\begin{equation}
    I(T)  =I_t \; e^{-k_1 (T-t)}+\theta_1 \left(1-e^{-k_1 (T-t)}\right)+\beta_1 \int_t^T e^{-k(T-u)} d W(u) 
\end{equation}}
Therefore, the statistics of $I(T)$ (and similarly $Q(T)$) are given by 
{\small
\begin{equation}
\begin{aligned}
\label{expectationI}
E\left[I(T) \right] & = I_t \; e^{-k_1 (T-t)}+\theta_1 \left(1-e^{-k_1 (T-t)}\right) \stackrel{T \rightarrow \infty}{\longrightarrow} \theta_1 \\
Var\left[I(T)\right] & =\frac{\beta^2_1}{2 k_1}\left(1-e^{-2 k_1 (T-t)}\right) \stackrel{T \rightarrow \infty}{\longrightarrow} \frac{\beta^2_1}{2 k_1},
\end{aligned}
\end{equation}}
and the autocorrelation of $I$ can be derived as follows
{\footnotesize
\begin{equation}
\begin{aligned}
\label{CII}
&C_{I I}(s, s+\Delta s)\\&= E\left[\beta^2_1 \int_t^s e^{-k_1(s-u)} d W(u) \int_t^{s+\Delta s} e^{-k_1(s+\Delta s-u)} dW(u) \right] \\
& =\beta^2_1 e^{-k_1 s}  e^{-k_1(s+\Delta s)} E\left[\int_t^s e^{k_1 u} dW(u) \int_t^{s+\Delta s} e^{k_1 u} d W(u) \right] \\
&=\beta^2_1 e^{-2 k_1 s-k_1 \Delta s} \int_t^s e^{2 k_1 u} d u \\
& =\frac{\beta^2_1}{2 k_1} e^{-k_1 \Delta s}\left(1-e^{-2 k_1 s}\right) \stackrel{s \rightarrow \infty}{\rightarrow} \frac{\beta^2_1}{2 k_1} e^{-k_1 \Delta s}.
\end{aligned}
\end{equation}}

By solving the SDE in (\ref{dynamicsRbar}), we have 
{\footnotesize
\begin{equation}
\begin{aligned}
& d \bar{R}(s)=B\left(\sigma^2-\bar{R}(s)\right) d s+\sigma \sqrt{2 B \bar{R}(s)} d W(s) \\
& e^{B s} d \bar{R}(s)+B e^{B s} \bar{R}(s) d s=B \sigma^2 e^{B s} d s+\sigma \sqrt{2 B \bar{R}(s)} e^{B s} d W(s) \\
& d\left(e^{B s} \bar{R}(s)\right)=B \sigma^2 e^{B s} d s+\sigma \sqrt{2 B\bar{R}(s)} e^{B s} d W(s) \\
& e^{B T} \bar{R}(T)-\bar{R}(0)=B \sigma^2 \int_0^T e^{B s} d s+\sigma \int_0^T \sqrt{2 B \bar{R}(s)} e^{B s} d W(s) \\
&=\sigma^2\left(e^{B T}-1\right)+\sigma \int_0^T \sqrt{2 B \bar{R}(s)} e^{B s} d W(s).
\end{aligned}
\end{equation}}
\normalsize 
Using the properties of Ito integrals,  we can write the expectation, the variance, and the autocorrelation of $\bar{R}(T)$ as follows:

{\footnotesize
\begin{equation}
\begin{aligned}
&E[\bar{R}(T)]=e^{-B T} \bar{R}(0)+\sigma^2 \left(1-e^{-B T}\right) \stackrel{T \rightarrow \infty}{\longrightarrow} \sigma^2
\end{aligned}
\end{equation}}
{\small
\begin{equation}
\begin{aligned}
&  \operatorname{Var}(\bar{R}(T))=\sigma^2 E\left[\left(\int_0^T \sqrt{2 B \bar{R}(s)} e^{-B(T-s)}dW(s)\right)^2\right] \\
& =\sigma^2 \int_0^T E\left[2 B \bar{R}(s) e^{-2 B(T-s)}\right] d s \\
& =2 B \sigma^2 e^{-2 B T} \int_0^T\left[e^{-B s} \bar{R}(0)+\sigma^2\left(1-e^{-B s}\right)\right] e^{2 B s} d s \\
& =\sigma^2\left[2\left(\bar{R}(0)-\sigma^2\right)\left(e^{-B T}-e^{-2 B T}\right)+\sigma^2\left(1-e^{-2 B T}\right)\right] \\
& \stackrel{T \rightarrow \infty}{\longrightarrow} \sigma^4 \\
\end{aligned}
\end{equation}}

{\footnotesize
\begin{equation}
\begin{aligned}
\label{autocorr}
& C_{\bar{R} \bar{R}}(t, t+\Delta t)=\sigma^2 \mathbb{E}\left[\int_0^t \sqrt{2 B \bar{R}(s)} e^{-B\left(t-s\right)} dW(s) \right. \\ & \left. \quad \quad \quad \quad \quad \quad \quad \quad \quad \quad \int_0^{t+\Delta t}  \sqrt{2 B \bar{R}\left(s\right)}  e^{-B(t+\Delta t -s)} dW(s) \right] \\
& =\sigma^2 \int_0^t \mathbb{E}[2 B \bar{R}(s)] e^{-2 B(t-s)} e^{-B \Delta t} d s \\
& =2 B \sigma^2 e^{-2 B t} e^{-B \Delta t} \int_0^t e^{B s} \bar{R}(0)+\sigma^2\left(1-e^{-B s}\right) e^{2 B s} ds \\
& =\sigma^2 e^{-B \Delta t}\left(2\left(\bar{R}(0)-\sigma^2\right)\left(e^{-B t}-e^{-2B t}\right)+\sigma^2\left(1-e^{-2B t}\right)\right) \\
& \stackrel{t \rightarrow \infty}{\rightarrow} \sigma^4 e^{-B \Delta t}.
\end{aligned}
\end{equation}}
\section{Proof of Lemma~\ref{MP}} \label{lemmaMP}
We let $g: \mathbbm{R}^{\bar{d}} \rightarrow \mathbb{R}$,
and $$
\bar{e}(t, s)=E\left[g\left(\bar{S}^{x_0}(T)\right) \mid \bar{S}^{x_0}(t)=s\right].
$$
We calculated the weak approximation error as 
\begin{equation}
\begin{aligned}
\epsilon_T&=E\left[g\left(P X(T)\right) \mid X(0)=x_0\right]\\&-E\left[g\left(\bar{S}^{x_0}(T)\right) \mid \bar{S}^{x_0}(0)=P x_0\right]
 \\& =E\left[\bar{e}\left(T, S(T)\right) \mid X(0)=x_0\right]-\bar{e}\left(0, S(0)\right).
\end{aligned}
\end{equation}
Using Ito's formula, 
\begin{equation}
\begin{aligned}
    d \bar{e}\left(t, Px\right)&=\left(\partial_t \bar{e}\left(t, Px\right)+[\mathcal{G} \bar{e}](t,  x) \right)dt+\\& P b(t, x) \partial_s \bar{e}\left(t, P x\right) dW(t),
    \end{aligned}
\end{equation}
where,

{\small
\begin{equation}
[\mathcal{G} \bar{e}](t, x)=P a(t, x) \partial_s \bar{e}\left(t, P x\right)+\frac{\left(P b b^T P^T\right)(t, x)}{2} \partial_{s s} \bar{e}\left(t, P x\right).
\end{equation}}
Therefore,
\begin{equation}
\begin{aligned}
\epsilon_T&=E\left[\int_0^T \left(\partial_t \bar{e}\left(t, S(t)\right)+[\mathcal{G} \bar{e}](t, X(t))\right) dt\right].
\end{aligned}
\end{equation}

In contrast, $\bar{e}(t, x)$ is the solution of the projected KBE,
$$
 \left\{\begin{aligned}
&\partial_t \bar{e}(t, x)+[\mathcal{\overline{\mathcal{G}}} \bar{e}](t, x)=0, \; t<T \\&
\bar{e}(T, \cdot) =g(\cdot),
\end{aligned} \right.
$$
where $[\mathcal{\overline{\mathcal{G}}} \bar{e}]=\bar{a}^{\left(x_0\right)} \partial_s \bar{e}+\frac{\bar{b} \bar{b}^{T^{\left(x_0\right)}}}{2} \partial_{s s} \bar{e}$. Consequently, 
{\small
$$
\begin{aligned}
& \epsilon_T=\int_0^T E\left[\left(P a(t, X(t))-\bar{a}^{\left(x_0\right)}\left(t, P X(t)\right)\right) \partial_s \bar{e}\left(t, P X(t)\right)\right] d t \\
& +\frac{1}{2} \int_0^T E\left[\left(\left(P b b^T P^T\right)(t, X(t))-\left(\bar{b} \bar{b}^T\right)^{\left(x_0\right)}\left(t, P X(t)\right)\right)\right. \\& \; \; \; \; \; \times \partial_{s s} \bar{e}\left(t, P X(t)\right) \left. \right] d t.
\end{aligned}
$$}

We chose $\bar{a}^{\left(x_0\right)}$ and $\bar{b}^{\left(x_0\right)}$ such that $\epsilon_T=0$. Using the tower property,

{\footnotesize
$$
\begin{aligned}
& E\left[\left(P a(t, X(t))-\bar{a}^{\left(x_0\right)}\left(t, P X(t)\right)\right) \partial_s \bar{e}\left(t, P X(t)\right)\right] \\
& =E\left[E\left[\left(P a(t, X(t))-\bar{a}^{\left(x_0\right)}\left(t, P X(t)\right)\right) \partial_s \bar{e}\left(t, P X(t)\right) \mid P X(t)\right]\right] \\
& =E\left[\left(E\left[\left(P a\right)(t, X(t)) \mid P X(t)\right]-\bar{a}^{\left(x_0\right)}\left(t, P X(t)\right)\right) \partial_s \bar{e}\left(t, P X(t)\right)\right].
\end{aligned}
$$}

\vspace{2mm}
Using a similar derivation for the second term in the expression of $\epsilon_T$, the expressions of $\bar{a}^{\left(x_0\right)}$ and $\bar{b}^{\left(x_0\right)}$ in (\ref{newdriftdiffusion}) ensure that $\epsilon_T=0$ for all bounded continuous functions $g$. Therefore, for all fixed $T$ and initial conditions $x_0$, $\left.S(t)\right|_{\left\{X(0)=x_0\right\}}=\left.P X(t)\right|_{\left\{X(0)=x_0\right\}}$ and $\bar{S}^{\left(x_0\right)}(t)$ have the same conditional distribution for all $t \in[0, T]$.
\section{Proof of Proposition~\ref{prop1}} \label{rice}
We write the conditional PDF $f_{I(s) \mid R(s)}(x \mid r)$ for $r>x^2$ as follows:
\begin{equation}
    f_{I(s) \mid R(s)}(x \mid r)=\frac{f_{I(s) R(s)}(x,r)}{f_{R(s)}(r)}=\frac{f_{R(s) \mid I(s)}(x,r) f_{I(s)}(x)}{f_{R(s)}(r)}.
\end{equation}
First, we know that $\sqrt{R(s)}$ has a Rice distribution with parameters $(\sqrt{2}m(s), \sigma(s))$. The PDF of $\sqrt{R(s)}$, for $ x \geq 0$, is given by 
\begin{equation}
    f_{\sqrt{R(s)}}(x)=\frac{x}{\sigma^2(s)} \exp\left(-\frac{2 m^2(s)+x^2}{2 \sigma^2(s)}\right) \mathcal{
I}_0\left(\frac{x\sqrt{2} m(s)}{\sigma^2(s)}\right).
\end{equation}
\\
Therefore, for $ r \geq 0$,
\begin{equation}
\label{pdfR1}
\begin{aligned}
f_{R(s)}(r)&=\frac{1}{2 \sqrt{r}} f_{\sqrt{R(s)}}(\sqrt{r}) \\& =\frac{1}{2 \sigma^2(s)} \exp\left(-\frac{2 m^2(s)+r}{2 \sigma^2(s)}\right) \mathcal{
I}_0\left(\frac{m(s) \sqrt{2 r}}{\sigma^2(s)}\right).
\end{aligned}
\end{equation}
In contrast, 
\begin{equation}
    f_{R(s) \mid I(s)}(r \mid x)=f_{I^2(s)+Q^2(s) \mid I(s)=x}(r)=f_{Q^2(s)}(r-x^2)
\end{equation}
We know that $\frac{Q^2(s)}{\sigma^2(s)}$ has a noncentral chi-square distribution with one degree of freedom and the parameter $\lambda(s)=\frac{m^2(s)}{\sigma^2(s)}$. Therefore, for $r>x^2$,
{\footnotesize
\begin{equation}
\begin{aligned}
    &f_{R(s) \mid I(s)}(r \mid x) \\& =\frac{1}{2 \sigma^2(s)} \exp\left(-\frac{r-x^2}{2 \sigma^2(s)}-\frac{\lambda(s)}{2}\right)\left(\frac{r-x^2}{\lambda(s) \sigma^2(s)}\right)^{-\frac{1}{4}} \\& \times \mathcal{
I}_{-\frac{1}{2}}\left( \frac{\sqrt{\lambda(s) (r-x^2)}}{\sigma(s)}\right).
\end{aligned}
\end{equation}}
Finally, for $ r>x^2$, we obtain 
{\footnotesize
\begin{equation}
f_{I(s) \mid R(s)}(x \mid r)=\frac{\exp\left(\frac{x m(s)}{\sigma^2(s)}\right) \left(\frac{r-x^2}{m^2(s)}\right)^{-\frac{1}{4}}\mathcal{
I}_{-\frac{1}{2}}\left( \frac{m(s)\sqrt{ (r-x^2)}}{\sigma^2(s)}\right)}{\sqrt{2 \pi} \sigma (s) \mathcal{
I}_0\left(\frac{m(s) \sqrt{2 r}}{\sigma^2(s)}\right)}.
\end{equation}}
\section{Proof of Proposition~\ref{prop2}} \label{hoyt}
We write the conditional PDF $f_{I^2(s) \mid R(s)}(x \mid r)$ for $r \geq x$ as follows:
{\small
\begin{equation}
\label{pdf}
    f_{I^2(s) \mid R(s)}(x \mid r)=\frac{f_{I^2(s) R(s)}(x,r)}{f_R(s)(r)}=\frac{f_{R(s) \mid I^2(s)}(x,r) f_{I^2(s)}(x)}{f_R(r)}.
\end{equation}}
Additionally, $R(s)$ has a square Hoyt distribution, and its PDF can be written as follows \cite{romero2017new}:
\begin{equation}
\begin{aligned}
\label{pdfR}
f_{R(s)}(r)&=\frac{1}{2 \sigma_1(s) \sigma_2(s)} \exp\left(- \frac{r(\sigma_1^2(s)+\sigma_2^2(s))}{4 \sigma^2_1(s) \sigma^2_2(s)}\right) \\& \times  \mathcal{
I}_0\left(\frac{r (\sigma_1^2(s)-\sigma_2^2(s)) }{4 \sigma^2_1(s) \sigma^2_2(s)}\right), \; r \geq 0.
\end{aligned}
\end{equation}
\\ \\
We can express $f_{R(s) \mid I^2(s)}(x,r)$ and $f_{I^2(s)}(x)$ as follows because $\frac{I^2(s)}{\sigma^2_1(s)}$ and $\frac{Q^2(s)}{\sigma^2_2(s)}$ follow chi-squared distributions with one degree of freedom:
\\
{\small
\begin{equation}
\label{pdf2}
   f_{R(s) \mid I^2(s)}(x,r)=\frac{1}{\sqrt{2 \pi}\sigma^2_2(s)} \left(\frac{r-x}{\sigma^2_2(s)}\right)^{-\frac{1}{2}} \exp\left(-\frac{r-x}{2 \sigma^2_2(s)}\right),
\end{equation}}
and 
{\small
\begin{equation}
\label{pdf3}
   f_{I^2(s)}(x)=\frac{1}{\sqrt{2 \pi}\sigma^2_1(s)} \left(\frac{x}{\sigma^2_1(s)}\right)^{-\frac{1}{2}} \exp\left(-\frac{x}{2 \sigma^2_1(s)}\right).
\end{equation}}
\\
Substituting (\ref{pdfR}), (\ref{pdf2}), and (\ref{pdf3}) into (\ref{pdf}), we obtain:
{\small
\begin{equation}
    f_{I^2(s) \mid R(s)}(x \mid r)=\frac{\left(r-x\right)^{-\frac{1}{2}} \exp\left(-\frac{r-x}{2 \sigma^2_2(s)}-\frac{x}{2 \sigma^2_1(s)}\right) x^{-\frac{1}{2}}}{\pi \exp\left(- \frac{r(\sigma_1^2(s)+\sigma_2^2(s))}{4 \sigma^2_1(s) \sigma^2_2(s)}\right) \mathcal{
I}_0\left(\frac{r (\sigma_1^2(s)-\sigma_2^2(s)) }{4 \sigma^2_1 (s)\sigma^2_2(s)}\right).}
\end{equation}}
\section{KBE Solver} \label{KBEsolver}
To reduce the number of parameters in (\ref{PDE}), we introduced $\tilde{z}=w-z$, $\Tilde{Z}(T)=w-\bar{Z}(T)=\tilde{z} -\int_t^T \mathbbm{1}_{\{\bar{R}(s)<\gamma^2\}} ds$, and $\tilde{g}(x)=\mathbbm{1}_{\{x<0\}}$. We let $\tilde{v}(t, x,\tilde{z}) = \mathbb{E}[\tilde{g}(\Tilde{Z}(T)) \mid \bar{R}(t)=x, \Tilde{Z}(t)=\tilde{z}]$. Then, $\tilde{v}$ solves the following PDE, for $x \geq 0,\; 0 \leq t\leq T, \; \text{and} \;  -T\leq \tilde{z} \leq T$:
\begin{equation}
\label{PDE2}
\left\{\begin{aligned}
&\partial_t {\tilde{v}}(t,x,\tilde{z})+B\left(\sigma^2-x\right) \partial_x {\tilde{v}}(t,x,\tilde{z})-\bar{c}(\tilde{z}) \partial_{\tilde{z}} {\tilde{v}}(t,x,\tilde{z}) \\ &+\sigma^2 B x \partial_{x x}{\tilde{v}}(t,x,\tilde{z})=0, \\
&{\tilde{v}}(T, x,\tilde{z})=\tilde{g}(\tilde{z}), \quad \text{(+ Boundary conditions)}.
\end{aligned}\right.
\end{equation}
Instead of solving the PDE (\ref{PDE}) again for each value of $w$, we solve (\ref{PDE2}) just one time, and for a fixed $w$, we take $v(t, x, z) = \tilde{v}(t, x, w - z)$.

To begin, we must establish the boundary conditions for (\ref{PDE2}). If $\tilde{z}<0$, then $\Tilde{Z}(T)<0$. Consequently, $\tilde{v}(t, x,\tilde{z})=P(\Tilde{Z}(T)<0| \bar{R}(t)=x, \Tilde{Z}(t)=\tilde{z})=1$, implying that it is only necessary to solve the PDE for $0\leq \tilde{z} \leq T$. Additionally, 
\begin{itemize}
   \item If $\tilde{z}=0$, $\tilde{v}(t, x,0)=1$ for $t< T$, and $\tilde{v}(T, x,0)=0$.
    \item If $\tilde{z}=T$, $\tilde{v}(t, x,T)=0$ for $0 \leq t\leq T$.
\end{itemize}
Notably, $\tilde{Z}(T)=\tilde{z} -\int_t^T \mathbbm{1}_{\{\bar{R}(s)<\gamma^2\}} ds \geq \tilde{z} -(T-t)$, which means that if $\tilde{z} -(T-t) \geq 0$, then $\tilde{Z}(T) \geq 0$, and $\tilde{v}(t, x,\tilde{z} )=0$. Therefore, the boundary conditions for $\tilde{z}$ are as follows:
\begin{itemize}
    \item If $\tilde{z}=0:$ $\tilde{v}(t, x,0)=1$ for $t<T$, and  $\tilde{v}(T, x,0)=1$, 
    \item If $\tilde{z} \geq T-t :$ $\tilde{v}(t, x,\tilde{z})=0$ for $0 \leq t\leq T$.
\end{itemize}
To discretize the PDE in (\ref{PDE2}) using finite differences, we introduced the following grids:
$$
\begin{aligned}
& 0=t_0<t_1<\ldots<t_{N_t+1}=T, \quad  \\
& 0=x_0<x_1<\ldots<x_{N_x+1}=x_b, \text { and } \\
& 0=z_0<z_1<\ldots<z_{N_z+1}=T,
\end{aligned}
$$
where $ x_b \in \mathbb{R}$ is a sufficiently large constant. \\
For the space dimension $x$, we approximated the unbounded domain $[0, \infty )$ using a suitably selected interval $\left[0, x_b\right]$. For simplicity, we took $N_z=N_t$ and used uniform grids in all dimensions, obtaining 
$$
\begin{aligned}
&t_n=n \Delta t, \quad n=0, \ldots, N_t,\\
&x_i=i \Delta x, \quad i=0, \ldots, N_x, \text{ and}\\
&z_j=j \Delta t, \quad j=0, \ldots, N_t.
\end{aligned}
$$ 
We let $v_{ij}^n \equiv \tilde{v}\left(t_n, x_i, z_j \right) \equiv \tilde{v}\left(n \Delta t,i \Delta x, j \Delta t\right)$ and employ the central differences for discretization in both the $x$ and $z$ dimensions, as follows,
$$
\begin{aligned}
& \left.\frac{\partial \tilde{v}}{\partial x}\right|_{ij} ^n=\frac{v_{i+1 j}^n-v_{i-1 j}^n}{2 \Delta x}+\mathcal{O}\left(\Delta x^2\right), \\
& \left.\frac{\partial \tilde{v}}{\partial x^2}\right|_{ij} ^n=\frac{v_{i+1 j}^n-2 v_{ij}^n+v_{i-1 j}^n}{\Delta x^2}+\mathcal{O}\left(\Delta x^2\right),
\\ & \left.\frac{ \partial \tilde{v}}{\partial \tilde{z}}\right|_{ij} ^n=\frac{v_{i j+1}^n-v_{i j-1}^n}{2 \Delta t}+\mathcal{O}\left(\Delta t^2\right). \\
\end{aligned}
$$
From (\ref{PDE2}), $\partial_t {\tilde{v}}=-\bar{a}(t,x)\partial_x {\tilde{u}}+\bar{c}(\tilde{z}) \partial_{\tilde{z}} {\tilde{v}}-  \frac{\bar{b}(t,x)^2}{2}\partial_{x x}{\tilde{v}}$. Then, using the previous discretizations, we obtain 

\begin{equation}
\begin{aligned}
    \partial_t {\tilde{v}}|_{ij} ^n =& -\underbrace{ \left(\frac{\bar{a}(t_n,x_i)}{2 \Delta x}+\frac{\bar{b}(t_n,x_i)^2}{2 \Delta x^2}\right)}_{=\alpha_{u,i}} v_{i+1 j}^n +\underbrace{\frac{\bar{c}(x_i)}{2 \Delta t}}_{=\gamma_i} v_{ij+1}^n\\ & +\underbrace{\left(\frac{\bar{a}(t_n,x_i)}{2 \Delta x}-\frac{\bar{b}(t_n,x_i)^2}{2 \Delta x^2}\right)}_{=\alpha_{l,i}} v_{i-1 j}^n -\frac{\bar{c}(x_i)}{2 \Delta t} v_{i j-1}^n  \\&  -\underbrace{\frac{\bar{b}(t_n,x_i)^2}{\Delta x^2}}_{=\beta_i} v_{i j}^n +\mathcal{O}\left(\Delta t^2\right)+\mathcal{O}\left(\Delta x^2\right)
     \\=& -\alpha_{u,i} v_{i+1 j}^n +\alpha_{l,i} v_{i-1 j}^n +\gamma_i v_{ij+1}^n -\gamma_i  v_{i j-1}^n+\beta_i v_{i j}^n \\& +\mathcal{O}\left(\Delta t^2\right)+\mathcal{O}\left(\Delta x^2\right).
\end{aligned}
\end{equation}

To verify the boundary conditions for $\tilde{z}$, $v_{ij}^n$ should satisfy the following:
\begin{itemize}
    \item $v_{i0}^n=1$, if $n<N_t+1$, and $v_{i0}^{N_t+1}=0$,
    \item $v_{ij}^n=0$, if $j \geq N_t+1-n$ for $0 \leq n\leq N_t+1$.
\end{itemize}
For the space dimension $x$, we implement nonreflective boundary conditions at $x=0$ and $x=x_b$ (i.e., $v_{N_x+1j}^n$ and $v_{0j}^n$ are extrapolated from known values):
\begin{itemize}
    \item $v_{0j}^n \approx 2 v_{1j}^n-v_{2j}^n$,
    \item $v_{N_x+1 j}^n \approx 2 v_{N_x j}^n-v_{N_x-1 j}^n.$
\end{itemize}
For each time step, the vector of unknowns is written as 
$$\mathbf{v}^n \equiv\left(v_{11}^n, \ldots, v_{N_x1}^n, \ldots \; \ldots, v_{1N_t}^n, \ldots, v_{N_xN_t}^n \right)^T.$$
We formulated the problem as a linear system for each time step:
$$
\left(\left.\frac{\partial \tilde{v}}{\partial t}\right|_{ij}^n\right)_{i=1,\ldots,N_x\atop j=1,\ldots,N_t}
=\mathbf{G}^n \mathbf{u}^n+ \mathbf{W}^n +\mathcal{O}\left(\Delta x^2\right)+\mathcal{O}\left(\Delta t^2\right),
$$
where $\mathbf{G}^n \in \mathbb{R}^{N_tN_x \times N_tN_x }$ and $\mathbf{W}^n \in \mathbb{R}^{N_tN_x}$  are the discretization operators at time $t_n$, including the boundary conditions, and
\begin{equation}
\mathbf{G}^0=     
\begin{bmatrix}
\mathbf{A}& \mathbf{\Gamma} & 0 & \cdots & 0 \\
-\mathbf{\Gamma} & \mathbf{A} & \mathbf{\Gamma} & \cdots & 0 \\
0 & -\mathbf{\Gamma} & \mathbf{A} & \cdots & 0 \\
\vdots & \vdots & \vdots & \ddots & \vdots \\
0 & 0 & 0 & -\mathbf{\Gamma} & \mathbf{A} \\
\end{bmatrix}
\end{equation}
where 
$\mathbf{A}\in \mathbb{R}^{N_x \times N_x }$ is given by

{\footnotesize
$$\mathbf{A}=\left(\begin{array}{cccccc}
a_{11} &a_{12} & 0 & 0 & \cdots & 0 \\
\alpha_{l, 2} & \beta_2 & -\alpha_{u,2} & 0 & \cdots & 0 \\
0 & \ddots & \ddots & \ddots & \ddots & \vdots \\
\vdots & \ddots & \ddots & \ddots & \ddots & 0 \\
\vdots & \ddots & \ddots & \alpha_{l, N_x-1} & \beta_{N_x-1}& -\alpha_{u, N_x-1} \\
0 & 0 & \cdots & 0 & a_{N_x N_x -1}& a_{N_x N_x}

\end{array}\right),
$$}
where $a_{11}=2 \alpha_{l,1}+\beta_1, \; a_{12}=-(\alpha_{l,1}+\alpha_{u,1}), \;  a_{N_x N_x -1}=\alpha_{l,N_x}+\alpha_{u,N_x}, \; a_{N_x N_x }= -2 \alpha_{u,N_x}+\beta_{N_x}$. \\
$\mathbf{\Gamma}\in \mathbb{R}^{N_x \times N_x }$ is given by 
\begin{equation}
\mathbf{\Gamma} = 
\begin{bmatrix}
\gamma_1 & 0 & \cdots & 0 \\
0 & \gamma_2 & \cdots & 0 \\
\vdots & \vdots & \ddots & \vdots \\
0 & 0 & \cdots & \gamma_{N_x} \\
\end{bmatrix}
\end{equation}
We revisit the boundary condition $v_{ij}^n=0$, if $j \geq N_t+1-n$, for $0 \leq n\leq N_t+1$. This condition introduces a time-step dependency in the matrix $\mathbf{G}^n$. In summary, matrix $\mathbf{G}^n$ is an evolved version of $\mathbf{G}^0$ characterized by the following transformation:
\begin{itemize}
    \item Take $\mathbf{G}^n=\mathbf{G}^0$,
    \item Make the last $n N_x$ rows of $\mathbf{G}^n$ equal to zeros.
    \item If $n<N_t$, replace $\mathbf{G}_{(N_t-n)N_x+k,(N_t-n-1)N_x+k}^n$ with $\gamma_k$ for $k=1, \cdots, N_x$.
    \item If $n<N_t$, replace $\mathbf{G}_{(N_t-n-1)N_x+k,(N_t-n)N_x+k}^n$ with zero for $k=1, \cdots, N_x$.

\end{itemize}
Using the boundary condition $v_{i0}^n=1$ if $n<N_t+1$, we can conclude that $\mathbf{W}^n=(\gamma_1,\gamma_2,\cdots, \gamma_{N_x},0,0,\cdots,0)^T$, for $n<N_t+1$.\\
We used the Crank--Nicholson scheme: 
\begin{equation}
    \begin{aligned}
\frac{\mathbf{v}^n -\mathbf{v}^{n-1} }{\Delta t}=\frac{1}{2} \left (\mathbf{G}^n \mathbf{v}^n+ \mathbf{W}^n \right)+\frac{1}{2} \left(\mathbf{G}^{n-1} \mathbf{v}^{n-1}+ \mathbf{W}^{n-1} \right).  
    \end{aligned}
\end{equation}
Hence, the update for the time step in the matrix-vector form is represented as follows: 
\begin{equation}
\label{backward}
    \begin{aligned}
\mathbf{v}^{n-1}&=\left(\mathbf{I}+ \frac{1}{2} \Delta t \mathbf{G}^{n-1}\right)^{-1}\left(\mathbf{I}-\frac{1}{2} \Delta t \mathbf{G}^n\right) \mathbf{v}^n  \\ &-\frac{1}{2}\Delta t \left(\mathbf{I}+\frac{1}{2}\Delta t \mathbf{G}^{n-1}\right)^{-1} \left(  \mathbf{W}^n+  \mathbf{W}^{n-1}\right).
    \end{aligned}
\end{equation}
We have the terminal time condition at $t=T$: $v_{ij}^{N_t+1}=\tilde{g}(z_j)$ for $i=1,\ldots,N_x$, $ j=1,\ldots,N_t$. Therefore, we can solve (\ref{backward}) starting from $\mathbf{v}^{N_t+1}$ until obtaining $\mathbf{v}^{0}$. Moreover, $\mathbf{G}^{N_t+1}=\mathbf{0_{N_tN_x \times N_tN_x}}$ and $\mathbf{W}^{N_t+1}=\mathbf{0_{N_t N_x}}$.
\begin{onehalfspacing}

	\addcontentsline{toc}{chapter}{References}
	\newcommand{\BIBdecl}{\setlength{\itemsep}{0pt}}
		\bibliographystyle{IEEEtran}
		\bibliography{References}
\end{onehalfspacing}
\end{document}